\documentclass[graybox]{svmult}

\usepackage{subfig}
\usepackage{url}
\newcommand{\mean}[1]{\overline{#1}}
\usepackage{xcolor}
\usepackage{type1cm}        % activate if the above 3 fonts are
\usepackage{rotating}
\usepackage{makeidx}         % allows index generation
\usepackage{graphicx}        % standard LaTeX graphics tool
\usepackage{multicol}        % used for the two-column index
\usepackage[bottom]{footmisc}% places footnotes at page bottom

\usepackage{newtxtext}       % 
\usepackage[varvw]{newtxmath}       % selects Times Roman as basic font

\usepackage{imakeidx}
\makeindex[options={-s svind.ist}]        
\usepackage{amsmath}
\usepackage{mathtools}
\usepackage{stackengine} 
\usepackage[bb=boondox]{mathalfa}
\newcommand{\of}[1]{\left(#1\right)}
\newcommand{\af}[1]{\left[#1\right]}
\newcommand{\uf}[1]{\left\{#1\right\}}

\newcommand{\skp}{\hspace{2pt}}
\newcommand{\skpp}{\hspace{1pt}}

\newcommand \stacksymbol[3]{\mathrel{\stackunder[2pt]{\stackon[4pt]{$#1$}{$\scriptscriptstyle#2$}}{$\scriptscriptstyle#3$}}}

\usepackage{lineno}
\usepackage{authblk}% for affiliations

\begin{document}
% Uncoment if line numbering is needed
%\linenumbers

\title*{Hybrid methods in reaction-diffusion equations}
\author{Tom\'as Alarc\'on\orcidID{0000-0002-8566-3676}, Natalia Bri\~nas-Pascual\orcidID{0009-0006-2974-3133}, Juan Calvo\orcidID{0000-0001-8459-8799}, Pilar Guerrero\orcidID{0000-0002-5522-7339} and Daria Stepanova\orcidID{0000-0003-1771-1987}}

%%% Daria: I put this in case other affiliations are not working, please add your affiliations correctly + the email of the corresponding author
%% If we add the full address, we should add it to all affiliations
%\affil[1]{Instituci\'{o} Catalana de Recerca i Estudis Avan\c{c}ats (ICREA), Passeig Llu\'{\i}s Companys, 12, 08010 Barcelona, Spain}%
%\affil[2]{Centre de Recerca Matem\`{a}tica (CRM), Campus de Bellaterra, Edifici C, 08193 Bellaterra (Barcelona), Spain}%
%\affil[3]{Departament de Matem\`{a}tiques, Universitat Aut\`{o}noma de Barcelona, Campus de Bellaterra, Edifici C, 08193 Bellaterra (Barcelona), Spain}%
%\affil[4]{Barcelona Collaboratorium for Predictive and Theoretical Biology, Barcelona, Spain}%
%\affil[5]{Grupo Interdisciplinar de Sistemas Complejos, Departamento de Matem\'aticas, Universidad Carlos III de Madrid, 28911 Legan\'es, Madrid, Spain}%
%\affil[6]{Departamento de Matem\'atica Aplicada and  Research Unit ``Modeling Nature'' (MNat), Avda. Fuentenueva s/n, Facultad de Ciencias, Universidad de Granada, 18071 Granada, Spain}%
%\affil[*]{Corresponding author, \texttt{talarcon@crm.cat}}
\institute{Tom\'as Alarc\'on \at Instituci\'o Catalana de Recerca i Estudis Avan\c cats, Passeig Llu\'{\i}s Companys, 12, 08010 Barcelona, Spain, Centre de Recerca Matem\`atica, Edifici C, Campus de Bellaterra, 08192 Cerdanyola del Vall\`es, Spain, Departament de Matem\`atiques, Universitat Aut\`onoma de Barcelona, Edifici C, Campus de Bellaterra, 08192 Cerdanyola del Vall\`es, Spain, Barcelona Collaboratorium for Predictive and Theoretical Biology, Wellington, 30, 08005 Barcelona, Spain \\ Corresponding author: \texttt{talarcon@crm.cat} 
\and 
Natalia Bri\~nas-Pascual and Pilar Guerrero \at  Grupo Interdisciplinar de Sistemas Complejos, Departamento de Matem\'aticas, Universidad Carlos III de Madrid, 28911 Legan\'es, Madrid, Spain
\and 
Juan Calvo \at Departamento de Matem\'atica Aplicada and Research Unit ``Modeling Nature'' (MNat), Avda. Fuentenueva s/n, Facultad de Ciencias, Universidad de Granada, 18071 Granada, Spain
\and 
 Daria Stepanova \at Centre de Recerca Matem\`atica, Edifici C, Campus de Bellaterra, 08193 Bellaterra, Barcelona, Spain, \email{dstepanova@crm.cat}
% \and 
% Name of Second Author \at Name, Address of Institute \email{name@email.address}
}
\maketitle

\abstract*{Each chapter should be preceded by an abstract (no more than 200 words) that summarizes the content. The abstract will appear \textit{online} at \url{www.SpringerLink.com} and be available with unrestricted access. This allows unregistered users to read the abstract as a teaser for the complete chapter.
Please use the 'starred' version of the \texttt{abstract} command for typesetting the text of the online abstracts (cf. source file of this chapter template \texttt{abstract}) and include them with the source files of your manuscript. Use the plain \texttt{abstract} command if the abstract is also to appear in the printed version of the book.}

\abstract{Simulation of stochastic spatially-extended systems is a challenging problem. The fundamental quantities in these models are individual entities such as molecules, cells, or animals, which move and react in a random manner. In big systems, accounting for each individual is inefficient. If the number of entities is large enough, random effects are negligible, and often partial differential equations (PDEs) are used in which the fluctuations are neglected. When the system is heterogeneous, so that the number of individuals is large in certain regions and small in others, the PDE description becomes inaccurate in certain regions. To overcome this problem, the so-called hybrid schemes have been proposed that couple a stochastic description in parts of the domain with its mean field limit in the others. In this chapter, we review the different formulations of this approach and our recent contributions to overcome several of the limitations of previous schemes, including the extension of the concept to multiscale models of cell populations.}

%\tableofcontents
\medskip 

\section{Introduction}
\label{sec:Intro}

Collective cell motion is ubiquitous in many biological phenomena, both in normal physiological processes as well as in pathological situations. For example, collective cell migration plays an essential role in vertebrate development. A system that is particularly amenable to both experimental and computational investigation is the neural crest. Neural crest cell populations are known to perform long-distance collective migration. If they do not reach the proper target or colonise an incorrect location, then improper cell differentiation or uncontrolled proliferation can ensue. This system has been studied extensively from a multidisciplinary perspective \cite{mclennan2015,giniunaite2020,martinson2023}. These investigations have shown that follow-the-leader behaviour, phenotype heterogeneity, and interactions with the extracellular microenvironment, specifically, the extracellular matrix, are essential to coordinate the movement of neural crest cells effectively. Angiogenesis, i.e. the growth of de novo vascular networks by sprouting from the existing vascular plexus, also requires coordinated cell migration. Similarly to cell migration in the neural crest, the formation of proper vascular patterns also requires phenotype heterogeneity and switching as well as interactions with the extracellular matrix \cite{stepanova2021}.

Pathological situations such as the growth of solid tumours also depend upon coordinated cell movement during invasion. This phenomenon demonstrates the emergence of cooperation between heterogeneous phenotypes to facilitate invasion within a dysregulated microenvironment, which is reminiscent of the collective cell motion occurring in development or wound-healing.

Collective cell motion, specifically invasion, has been modelled using several approaches, namely, individual-based models, stochastic models, and continuum models \cite{giniunaite2020}. In particular, to study invasion phenomena, a commonly used framework is that of reaction-diffusion systems (RD hereafter -both in the stochastic and deterministic versions). Within this framework, the collective behaviours associated with invasion are usually analysed in the context of travelling wave (TW) solutions. TWs are solutions where invasion takes place as a %sharp,
moving front that advances at constant velocity.

Reaction-diffusion models are used in fields as diverse as Physics, Biology, and Chemistry, to mention just a few. The fundamental entities in such models are individuals (molecules, animals, cells, etc), which move and react stochastically. For large systems, a purely individual-based approach becomes computationally inefficient. If the size of the system is large enough, one can use several techniques to derive macroscopic limits of the microscopic (individual-based) model. These limits often take the form of partial differential equation (PDE) models, in which the stochastic behaviour of individuals is replaced by an averaged version of the system, usually referred to as the mean-field limit. 

However, in many practical situations, the number of individuals is large in certain spatial regions and small in others, and the mean-field limit fails to capture the behaviour of the system in some regions having just a few individuals, whereas the stochastic or individual-based model may be inefficient in other regions where the number of individuals is huge. 

To overcome this problem, the so-called hybrid methodologies have been developed. Such methods combine the mean-field and stochastic descriptions of the system into a single scheme in such a way that one takes advantage of both the accuracy of the stochastic description and the efficiency of the mean-field limit. The basis of all such methods is to use the stochastic system in some parts of the domain and its mean-field limit, i.e. the corresponding discretised PDE model, in some other regions. Therefore, all versions of a hybrid methodology must address two key issues. First, the hybrid method must provide a criterion (or criteria) for, at each time, deciding which regions of the system will be treated as deterministic and which ones are considered stochastic. The second issue hybrid methods must tackle is the definition of the transition region (or interface) between deterministic and stochastic subdomains. A standard assumption is that the interface between the two domains is chosen so that both representations (mean-field and stochastic) describe the system with sufficient accuracy. Such a criterion ensures that the errors associated with the fluxes of individuals between the stochastic and the deterministic regions are small.

The paradigmatic example that has been used as a benchmark for most proposals of hybrid methods in RD systems is that of TWs in one spatial dimension (1D). TWs are particular solutions of RD systems in which a stable phase invades an unstable one. TWs take the form of fronts propagating at a constant speed (asymptotically). A specific example of an RD system that exhibits this type of solution is the Fisher-Kolmogorov equation (also known as the Fisher-Kolmogorov-Petrovski-Piscunov or FKPP equation -see e.g. \cite{Aronson1978} and references therein). The FKPP equation provides a suitable test case against which the performance of hybrid methods can be tested. The front of a given travelling wave represents a transition region between a domain at the back of the moving front, where the population described by the FKPP equation is large (close to its carrying capacity), and a region ahead of the front, where the population is much lower (in fact, close to zero, i.e. close to extinction). As a result, the latter region would be better suited for a stochastic description, whereas the former can be described using the mean-field version without significant loss of accuracy. The interface can be therefore located within the front region at a position that allows for both descriptions to be applicable. Of course, such an interface should advance at (roughly) the same speed as the front. 

Thus, the basic hybrid scheme should account for three key aspects, namely, (i) a criterion to decide which regions are mean-field and which ones are stochastic, (ii) appropriate boundary conditions at the interface so no artefacts are created by it, and, finally, (iii) a prescription for where the interface should be collocated so that it moves at the same rate as the fronts within the system. Several such proposals are summarised in section~\ref{sec:explore} below.

However, this basic scheme falls short in many problems of interest, namely, as soon as we consider more than one species.  
RD systems can have a much richer range of behaviours than TWs,  such as Turing-like patterns, standing waves, scroll waves, and more. 
Hybrid methods should be formulated so that they can cope with several interacting species and therefore account for these rich dynamics. In particular, hybrid methodologies should be able to account for more than one interface at once. 

Another limitation of the basic hybrid schemes is more fundamental. The reaction-diffusion Master Equation (RDME), frequently used to describe system dynamics in stochastic subdomains, is afflicted with a serious problem. Namely, it does not converge to its microscopic description in dimensions greater than one 
\cite{gillespie1976general}, even though a meaningful PDE exists in the limit of compartment size going to zero \cite{smith2016breakdown}. The reason for the breakdown of the RDME as the lattice size tends to zero is that the waiting time for bimolecular reactions to occur tends to infinity, thus leading to accuracy deterioration as a consequence of mesh refinement, whereby mesh size must be taken to be much greater than the reaction radius between molecules. Two strategies have been used to address this problem. One approach is to propose several \emph{generalised reaction-rates} to address the lower-bound issue \cite{hellander2016reaction, erban2009stochastic}. The second strategy is to formulate a convergent version of the Master Equation for reaction-diffusion systems. The convergent RDME (cRDME) is a general approach proposed by \cite{isaacson2013convergent}, whereby an approximation of Doi's model for the binary reactions, $A + B \rightarrow C$, is put forward. In this model, molecules within different voxels are allowed to interact through bimolecular reactions, which are allowed to occur between particles in voxels that lie within a predefined reaction radius, thus overcoming the artefact of vanishing reaction rates as voxel size tends to zero. Thus, to move the current schemes from one to a higher dimension, one needs to develop hybrid methods for the cRDME.

On top of all this, we should also keep in mind that population dynamics models are often multiscale models, where cells have intracellular dynamics that regulate behaviour as a function of external inputs or cues (e.g. proliferation regulated by nutrient availability). The intracellular state is sometimes accounted for by a structure variable. We should also formulate hybrid methods for stochastic multi-scale models of cell populations that extend the remit of existing hybrid methods for reaction-diffusion systems, by extending current efforts to structured (e.g. age-structured) populations.

 In this chapter, we review our recent efforts to try and fill these gaps. The rest of this Introduction is devoted to reviewing the landscape of hybrid methods of models for RD systems. Among these, the case of multiscale population models is specifically discussed in section~\ref{sec:age}. We then move on to summarise a hybrid method that accounts for several species and several interfaces, in section~\ref{sec:Multiple_species}. Next, in section~\ref{sec:hybrid_cRDME}, we formulate a hybrid method for the cRDME. 
 Finally, in the Discussion, we present our conclusions and elaborate on future directions.

\subsection{Hybrid Methods in a Nutshell}\label{sec:explore}

RD systems describe the behaviour of various chemical, physical, and biological phenomena, from the formation of spatial patterns in nonlinear chemical reactions to the dynamics of populations. These systems model the spatio-temporal evolution of the concentrations (densities) of interacting species over time and space.

Usually, when studying these systems, there exist different techniques which exploit the existence of different inherent scales. Scale separation can be with regard to time scales, spatial scales, or the size of different species, among others. 
There exist four different scales in modelling algorithms that are relevant to our discussion. The macroscopic scale is the coarsest scale, where the number of particles of the system is large enough that species abundance can be represented by densities. Within this regime, PDEs are used. Within the mesoscopic scale, particles are compartmentalised in different subdomains. Particle distribution within compartments is assumed to be homogeneous. The number of particles changes because of diffusion between compartments and reactions within compartments. We will assume that the mesoscopic dynamics is described by a continuous-time Markov process. The microscopic scale is a finer scale where particles are studied as individuals with continuous trajectories. An example of this dynamic is Brownian motion. In the case of cellular systems, there is an intracellular scale, where a high level of detail is used to describe sub-cellular processes, such as signalling pathways (e.g. the cell-cycle). 
Combining any two of these scales and their respective modelling tools into a hybrid model allows for the integration of the strengths of both methodologies.

The nature of reactions occurring in a biological system varies significantly and is influenced by factors such as the number of particles, the types of species involved, and other specific conditions. Chemical reactions in biological systems can be categorized as either fast or slow, based on the species involved and the rate of each reaction. This approach can be applied to reaction-diffusion systems by modelling diffusion deterministically and the remaining reactions stochastically, like the method described in \cite{Rossinelli2008}, where a hybrid tau-leaping method is employed to first simulate the stochastic reactions and then use a deterministic diffusion operator. Also, in \cite{Haseltine2002}, an adaptive partitioning strategy is implemented based on the fractional propensity functions for each reaction or diffusion jump event. High-frequency processes are simulated using deterministic equations and low-frequency processes by applying the Gillespie algorithm \cite{gillespie1976general}.  

\subsubsection{Spatially-Coupled Hybrid Methods}
\label{sec:hybrid_algorithm}
Systems that can be considered spatially homogeneous and have no further internal structure are typically amenable to simple simulation approaches. Hybrid methodologies can still play a role in cases where the evolution of very abundant species is coupled to much less abundant species, such as piecewise deterministic Markov processes \cite{davis1984,bressloff2017}. In general, more benefits are derived from the use of hybrid methods when we are describing systems with additional independent variables besides time. The most usual situation is that of systems with spatial degrees of freedom. 

For concreteness, consider the Fisher-Kolmogorov system and its travelling wave solution. In this regime, the number of particles ahead of the front is much smaller than the number of particles behind it, where they approach the carrying capacity. Therefore, while the deterministic, mean-field PDE describes the system accurately behind the front, fluctuations at, and more importantly, ahead of the front must be considered. This is achieved by dividing the domain into two subdomains: one where the model is treated stochastically due to the low number of particles, and another where a higher number of particles allows for a deterministic model. Note that in this specific situation, both subdomains are connected intervals. This need not be the case in more elaborate scenarios.

This approach of partitioning the domain into stochastic and deterministic subdomains has been implemented following several schemes and methodologies, which we review below. In general, hybrid methods are based on the idea of connecting the subdomains. Within the first region, populations are large enough to justify the use of a PDE. In the other region, stochastic algorithms such as the Gillespie algorithm must be applied. The main difference between the various hybrid methods lies in how the interface between regions is defined and the so-called interface conditions. A fixed interface can be considered, so that the boundaries between subregions do not change over time. Conversely, a dynamic interface can be defined, so the new boundaries of each subdomain are computed at each time step. Another characteristic to be considered is the part of the domain shared by both regions at the same time. They can be independent from one another or they can overlap. In the latter case, a third region must be considered where the stochastic and the deterministic algorithms coexist. See Table \ref{tab:different methods} for an account of several of the possibilities that are found in the literature.

\begin{table}[!t]
\caption{Different hybrid methods}
\label{tab:different methods}     
\begin{tabular}{p{2.4cm}p{3cm}p{3cm}p{3.5cm}}
\hline\noalign{\smallskip}
Reference & Domain Decomposition & Type of interface & Other characteristics \\
\noalign{\smallskip}\svhline\noalign{\smallskip}
Ferm et al. \cite{Ferm2010} & PDE + SSA + $\tau$-Leaping & Dynamic & Adaptive time steps\\
Flegg et al. \cite{flegg2012two} & Compartment-based model + Brownian dynamics & Fixed, no overlapping & Interface balanced flux \\

Franz et al. \cite{Franz2013} & PDE + Brownian dynamics  & Overlapping / empty intersection & Accurate simulation of mean and variance in the BD subdomain\\
Spill et al. \cite{spill2015hybrid}& PDE + Gillespie & Dynamic, overlapping & Adaptation for modelling competition of species \\
Harrison et al. \cite{harrison2016hybrid} & PDE + Compartment-based stochastic model & Dynamic, adaptive & Flux-based dynamics, adaptation for multiple species \\
Smith et al. \cite{smith2021growth} & PDE + Compartment-based stochastic model & Fixed, zero flux boundary condition & Deterministic domain growth \\
\noalign{\smallskip}\hline\noalign{\smallskip}
\end{tabular}

\end{table}

%#######################################################

The hybrid method builds upon these stochastic models by providing a computationally efficient framework that combines deterministic and stochastic regions, allowing for accurate simulation of travelling waves in complex environments.

We provide a generic description of a hybrid algorithm for a RD model. It is a basic version that relies on the default implementation of Gillespie's algorithm \cite{gillespie1976general}.

\paragraph{Generic Hybrid Algorithm}

\begin{enumerate}

    \item Set initial time and provide initial settings.
    \item Given a current time $t$, calculate the time to next event, $\tau$, via the Gillespie algorithm \cite{gillespie1976general}. This time is computed as
\begin{align}\label{eq:timeToNextEvent}
    \tau = -\frac{1}{\sum_ka_k}\log(r_1)
    \end{align}
    where $a_k$ denotes any of the non-zero propensity functions of the model and $r_1$ is a random number sampled from  a uniform distribution in the unit interval, i.e $r_1 \sim \mathcal{U}(0, 1)$.  
    \item Determine which event or reaction is happening at $t+\tau$. This is done by taking an independent second random number $r_2 \sim \mathcal{U}(0, 1)$ and looking for the event tag $l$ such that
    
\begin{align}\label{eq:whichStochasticEvent}
    \frac{\sum_{k=1}^{l-1}a_k}{\sum_{k}a_k}<r_2\le \frac{\sum_{k=1}^la_k}{\sum_{k}a_k}.
    \end{align}
    \item Advance up to the new time $t+\tau$ in the deterministic region via
    the corresponding partial differential equation.

    \item Update interface positions and overlapped regions.
    \item Return to step 2 and iterate until a prescribed final time is reached.

\end{enumerate}

Some points above need a more detailed explanation. Again, these are easily understood if we have in mind the situation of a single TW moving from left to right.

\paragraph{Interface Location}
The determination of the voxel(s) that constitute the interface region must be done on the basis that both the deterministic and the stochastic description should hold true at the same time. This is crucial, as interactions with neighbouring voxels will be both deterministic and stochastic. Therefore, we need to enforce large concentrations in those interface voxels. Their location is based on a trade-off between performance and accuracy, since the mean-field PDEs neglect fluctuations but are numerically cheaper. Thus, expanding the mean-field domain improves performance. For large enough populations, fluctuations usually scale with the square root of the number $N$ of individuals of a given species in a compartment $k$, $\sqrt{N(k)}$. A threshold $\Theta$ is set to classify each compartment as stochastic if ${N(k)} < \Theta$, or deterministic otherwise. 

\begin{figure}
    \centering
    \includegraphics[width=0.7\linewidth]{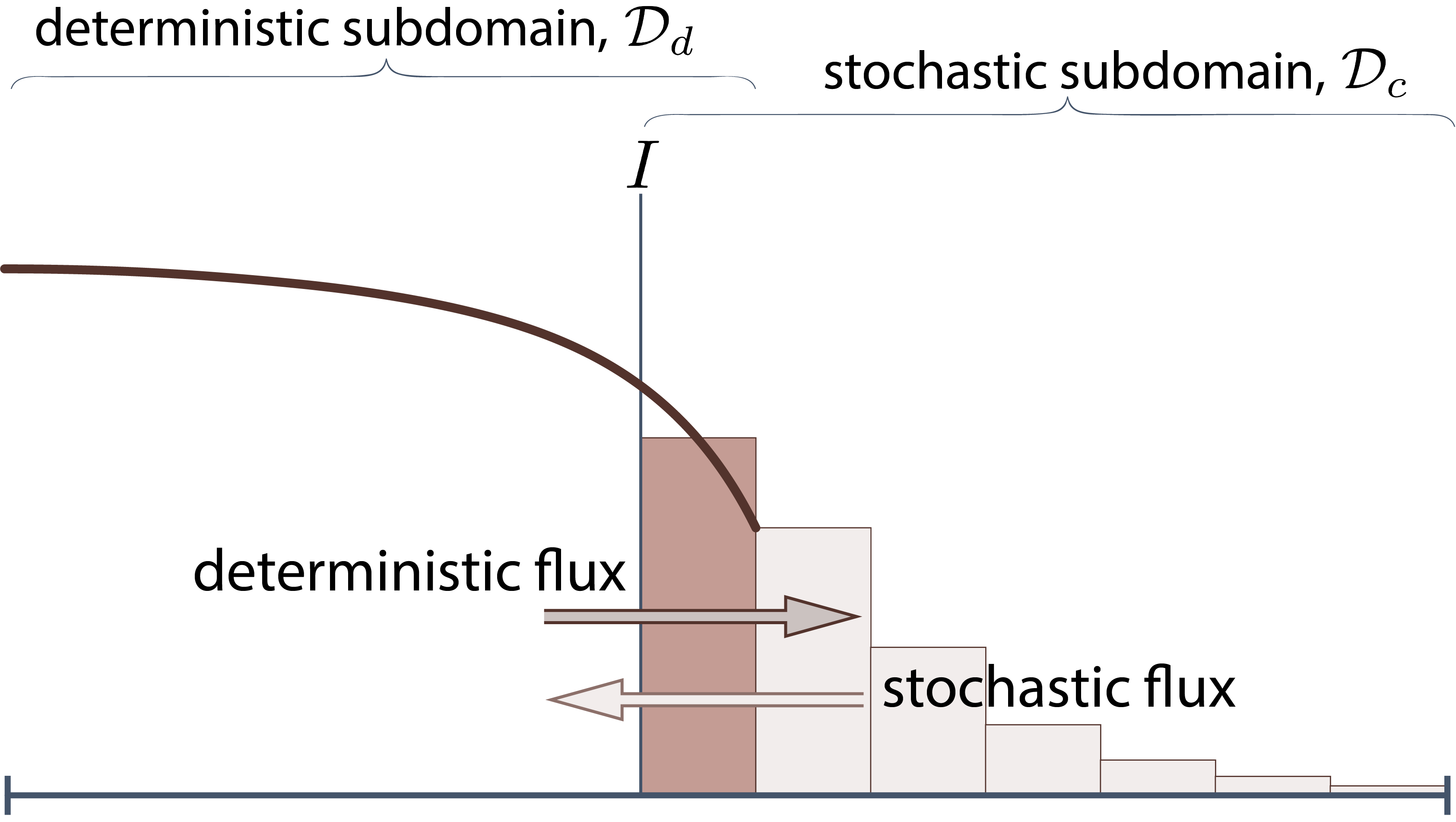}
    \caption{Schematic of a Hybrid Model Domain Division. The domain is divided into a deterministic domain and a stochastic domain  separated by an  interface (dark brown region). The deterministic flux and the stochastic flux indicate the transfer of particles across the interface, ensuring mass conservation and accurate coupling between the deterministic and the stochastic model.}
    \label{fig:hybrid regions}
\end{figure}

In the simple situation of TWs, assuming that the TW moves to the right, one sweeps the spatial domain from left to right and we place the interface at those voxels that are highly populated but to their right the concentration drops widely, as represented in Figure \ref{fig:hybrid regions}.

\paragraph{Fluxes and Reactions at Interfaces}
For simplicity, consider a single interface compartment, located at position $k_I$. Three different events contribute to the dynamics. Fluxes into and from the deterministic
compartment $(k_I -1)$ are modelled via the mean-field equations. Fluxes into and from the stochastic compartment
 $(k_I +1)$ and local reactions at $k_I$ are accounted for by transition rates in a stochastic way.

\paragraph{Moving the Interface}

As the number of particles for the species we are monitoring changes over time, the boundaries between the stochastic and deterministic regions need to be adjusted too, so the interfaces between them must be updated accordingly. This is done using the rules of interface placement described above.

Additionally, if the number of compartments in one region becomes too small, it may be more computationally efficient to merge this region with the other. Therefore, applying a minimum domain size condition helps to maintain efficiency by preventing very small regions.

Once the interface is properly allocated,  
we must ensure that all particle numbers in the stochastic domain are integer values and therefore density functions have to be updated too. 

\paragraph{Renormalisation of Particle Distribution}

When the interface changes its location, a compartment previously treated as deterministic can become part of the stochastic domain. Thus, after this is done we may get a particle number $N(k)$ which is not a natural number. We remedy this as follows. 
The fractional part 
\begin{equation}
 p = N(k) \mod(1)
\end{equation}
is interpreted as the probability that an additional particle is in this
compartment. By generating a random number $r \sim \mathcal{U}(0, 1)$  we can determine the allocation of the particle: if $r < p$ the particle is added to the box $k$; otherwise, it is placed in the mean-field domain. 
In the first case, the 
%number of particles
value $N(k)$ is increased by $(1-p)$, thus becoming an integer -the compartment densities are already scaled accordingly. 
In the second case, $N(k)$ is decreased by $p$, and the
densities are recomputed.

%%%%%%%%%%%%%%
\subsection{The Hybrid Method in the Context of Pulled Fronts Theory}

Deterministic models of pulled fronts describe the propagation of waves in reaction-diffusion systems where the leading edge of the wave determines the speed. These models are essential for understanding the minimal wave speed required for the invasion of homogeneous media. The classical example is the FKPP equation, for which we can predict the speed of pulled fronts based on the reaction and diffusion coefficients.  Namely, the wave speed $v^*$ of a pulled front is given by: 

$$v^*= 2 \sqrt{Dr},$$ 

\noindent where $D$ is the diffusion coefficient and $r$ is the growth rate. Recall that the FKPP equation was introduced to describe the spread of advantageous genes in a population. In this deterministic setting, this minimal wave speed is crucial for understanding how the advantageous trait invades the population.

In contrast, pushed fronts and waves in bistable media exhibit different propagation dynamics. Pushed fronts rely on the entire wave profile for propagation, while bistable waves require a threshold concentration for wave advancement.  The distinction between pulled and pushed fronts is crucial for understanding wave dynamics in RD systems. Pulled fronts are driven by the leading edge, while pushed fronts depend on the bulk of the wave. Bistable waves, on the other hand, require a critical concentration threshold to propagate. Hybrid methods for RD systems need to be adapted specifically to handle these distinct scenarios.

In many instances of applying hybrid methods to RD systems, the focus is on the dynamical description of the propagation of fronts into unstable states. In the deterministic description, the state of the system in the region far ahead of the front is linearly unstable. We can therefore distinguish between the so-called pulled and pushed fronts in more technical terms, as follows. For pulled fronts the asymptotic front speed (as time grows large) agrees with the linear spreading speed $v^*$ of infinitesimal perturbations around the unstable state -the perturbation ahead of the front spreads at speed $v^*$ and meanwhile pulls the rest of the front. Whereas for asymptotic front speeds strictly greater than $v^*$ we talk about pushed fronts because it is the nonlinear growth in the region behind the leading edge that controls the overall dynamics, as a truly nonlinear effect. In the classical FKPP equation, pushed fronts correspond to travelling wave solutions with a speed greater than $v^*$. In this particular case, we can relate these notions to the stability properties of those travelling solutions, as it is the wave with the slowest admissible speed, $v^*$, the one that attracts the dynamics for initial conditions that do not fall off too slowly at infinity including, in particular, compactly supported initial conditions, which are the relevant ones for our discussion here (see \cite{Aronson1978}). 

For more general FKPP equations of the form
\[
u_t=u_{xx}+f(u)
\]
where $f(0)=f(1)=0,\, f'(0)=1,\, f'(1)<0$, 
there is a result ensuring that when $f(u)/u\le f'(0)$ we will observe pulled fronts; this is to be expected, as under such conditions the nonlinearities will not be able to enhance growth and therefore we will not get pushed fronts. However, we lack a general mathematical framework which can predict whether, for a given RD equation, fronts will be pulled or pushed. The interested reader can have a look at \cite{Panja2004,vanSaarlos2003} and references therein for more information on these matters. To this point, we mention that in our discussion so far, the FKPP equation assumes a homogeneous medium, where the reaction and diffusion coefficients are constant; those restrictions can be lifted and as a consequence, we can get even richer dynamics (see \cite{Xin2000,Mendez2003}).

\paragraph{Stochastic and Heterogeneous Models of Pulled Fronts}

Deterministic pulled fronts often fail to capture the effects of demographic noise and heterogeneity present in biological systems. Extensive research has been conducted on stochastic and heterogeneous models of pulled fronts. These models incorporate fluctuations and spatial variability, offering a more accurate depiction of wave propagation in natural systems. Key studies include  work on large deviations and reaction-diffusion equations \cite{Freidlin1985}, and analysis of front speeds in heterogeneous media \cite{Mendez2003}.

It seems therefore natural to bypass the aforementioned limitations with the development of hybrid methods, as they integrate stochastic and deterministic descriptions.  In that regard, our focus is on stochastic descriptions given by discrete models; such as those reviewed in \cite{Panja2004} (see also section 7 in \cite{vanSaarlos2003}). There are also stochastic descriptions based on continuous points of view, such as stochastic partial differential equations with multiplicative noise. Although these are not directly related to the family of hybrid models we review in this contribution, the reader can have a look at, for example, \cite{Panja2004, Rocco2000, Xin2000} and references therein.
%we will reference some works along these lines in the sequel.

Stochastic models introduce fluctuations that can significantly alter the dynamics of wave propagation. For instance, pulled fronts in noisy environments can exhibit subdiffusive behaviour, where the root mean square (rms) wandering of the front scales as $t^{1/4}$
  rather than the $t^{1/2}$
  scaling seen in deterministic or pushed fronts \cite{Xin2000, Mendez2003}. Heterogeneous models further complicate this picture by introducing spatial variability in reaction rates and diffusion coefficients. Studies, such as those by \cite{Mendez2003}, explore how deterministic heterogeneities influence front speeds using singular perturbation analysis and Hamilton-Jacobi dynamics. These models show that front speeds, profile and location can vary significantly depending on the nature and scale of the heterogeneities, often requiring numerical simulations to capture the complex dynamics accurately \cite{Mendez2003, Xin2000}. This is critical in various applications, from chemical kinetics to biological invasions, where spatial variations can significantly impact the spread of fronts.
 
\paragraph{Application to Pulled Fronts}

Hybrid methods have primarily focused on pulled fronts due to their sensitivity to demographic noise at the leading edge. This is to be expected, as D. Panja notes \cite{Panja2004}: \textit{"It has to be noted that the majority of the studies on fluctuating fronts have been carried out on fluctuating ``pulled'' fronts (...). This is not very surprising -first of all, for the deterministic counterparts, it is the pulled fronts, for which there is a solid theoretical understanding. Secondly (and more importantly), it is the sensitivity of the pulled fronts to the dynamics of the leading edge, which is severely affected by the discreteness of particles and the lattice, or by the presence of external noise terms"}. Recall that for pushed fronts, their leading edges (which is the zone we would describe stochastically with a hybrid method) do not play any role in the front speed selection.

In the context of pulled fronts, hybrid methods have been particularly useful in studying the FKPP equation. The leading edge, characterized by low population densities, is treated stochastically, capturing the essential noise-driven dynamics. The bulk region, with higher densities, is modelled deterministically, reducing computational costs. This partitioning ensures that the front speed and shape are accurately simulated across the entire domain. However, applying these methods to bistable waves presents challenges, such as ensuring accurate interface placement and handling multiple stable states. Further research is needed to extend hybrid methodologies to these complex wave types.

Studies, such as those by \cite{Rocco2000,moro2004} demonstrate the utility of hybrid methods in capturing the subdiffusive wandering of pulled fronts in noisy environments \cite{Mendez2003}. These methods bridge the gap between purely deterministic and purely stochastic models, providing a more complete and computationally feasible approach to studying complex wave propagation in reaction-diffusion systems.

\subsection{Simulation Methods for Age-Structured Populations}
\label{sec:age}

In a nutshell, the dynamics of such complex biological systems as, for example, those mentioned in Introduction, is an emergent property of all the layers involved. It is in this rich context where hybrid models for structured, spatially extended populations can be extremely helpful, as the computational burden to keep track of every detail of such systems can be overwhelming. There have been a number of contributions extending the remits of the hybrid methodology to structured RD systems, among which we mention the work done by \cite{spill2015hybrid, delacruz2017coarse}.

Including structure variables is particularly relevant when dealing with cell populations, as it allows for the incorporation of demographic noise through an age structure. Other applications are, of course, possible. Our main focus in this area has been multiscale models of tumour growth. Here, hybrid methods are particularly helpful as they can cope with the usual demographic noise present in RD systems with a reasonable computational cost. Thus, we can construct multiscale models that can account for fluctuations in both the number of cells and the intracellular dynamics. For the latter, a cell cycle model is adopted, where the age structure provides the coupling with higher levels of description. 

The hybrid methodology put forward in \cite{spill2015hybrid, delacruz2017coarse} has a number of distinctive features: (i) The decomposition into deterministic and stochastic domains is performed only with respect to the spatial variable. (ii) A coarse-graining procedure is introduced, which integrates out the effects of the structure variable at those spatial regions where a deterministic description is used. The structure variable is fully taken into account in the stochastic regions, where the transition rates of the underlying birth-death-diffusion Markov process depends upon it. (iii) The dynamic adaptation of the spatial interface involves a renormalisation procedure (explained before in this review) that might demand the creation of new individuals. When this needs to be done, new ages are sampled from the equilibrium age distribution, which can be theoretically computed. The accuracy of this hybrid methodology is assessed by comparing it with full stochastic simulations, focusing on metrics such as the average position and speed of wavefronts \cite{delacruz2017coarse}. Regardless of how severe these simplifications might appear, our hybrid methodology can retrieve the travelling wave speeds that are obtained with a fully stochastic model, at a much lower computational cost. In a nutshell, this methodology is versatile and can be easily adapted to other scenarios of interest. Future research in this area might focus on refining these procedures to run them in higher spatial dimensions.

\section{Hybrid Method for Reaction-Diffusion Systems with Multiple Species}
\label{sec:Multiple_species}

Reaction-diffusion (RD) models are a flexible framework that can be used to model the interactions between several (spatially extended) species in a variety of settings such as ecological populations, chemicals, and cellular lineages, among others. To proceed with the hybrid methodology in this case, the spatial domain is decomposed into regions for each species. However, since multi-species RD systems produce heterogeneous patterns (e.g. Turing patterns, spatio-temporal oscillations, scroll waves, etc.), the decomposition might differ between the species that are present, and the chance arises that at a certain spatial region, we find interacting species that are described using different mathematical frameworks (i.e. mean-field or stochastic). 

Thus, the hybrid algorithm for multi-species RD systems involves multiple dynamic interfaces and key modifications to the standard setting to ensure efficient and accurate simulations. These adjustments are necessary to handle large and small stochastic time steps, deterministic interface reactions, and multiple interfaces for different species.

When the time to the next stochastic event, $\tau$, is bigger than the maximum time step allowed by the numerical method used for solving the mean-field PDE, $\tau_{PDE}$, several iterations of the finite difference scheme are required to evolve the PDE to the time $\tau$. During these iterations, the interface is treated as a Neumann no-flux boundary to prevent the PDE solution from leaking into the stochastic domain. This ensures that the deterministic and stochastic regions remain accurately separated during the simulation.

Conversely, if the Gillespie time step is smaller than the PDE convergence time step, the algorithm is modified so that the mean field domain is not iterated during every stochastic time step. Instead, it is iterated only after several stochastic steps, ensuring their cumulative time step is just under $\tau_{PDE}$. This adjustment helps to maintain computational efficiency without compromising the accuracy of the simulation.

Reactions at the interface can also be described deterministically by modifying the governing equation of the interface compartment. This approach replaces the stochastic treatment of interface reactions with a deterministic update equation. This modification ensures that the transition rates describing stochastic reactions in the interface compartment are set to zero, providing a more stable and accurate representation of interface dynamics.

The complexity of managing multiple species in hybrid models necessitates the introduction of multiple dynamic interfaces. Different species may occupy distinct regions with high concentrations, requiring separate interfaces for each species. The interface condition must be applied separately for each species. In regions where some species are modelled stochastically and others deterministically, reactions involving both types of species are handled stochastically. This ensures accurate coupling between the stochastic and deterministic domains for different species.

The hybrid model for multiple species introduces several key differences from the single-species hybrid model. In the initialisation phase, multiple species are defined, each with its own set of parameters such as diffusion coefficients, reaction rates, and thresholds. Unlike the single-species model, where a single stochastic and deterministic domain is managed, the multi-species model requires separate management of interfaces for each species. This includes handling multiple dynamic interfaces that can move independently based on the local population of each species.

Moreover, the multi-species hybrid model must account for coupled reactions involving species in different domains, ensuring accurate interactions between stochastic and deterministic regions. This increases the computational complexity of the model, requiring more efficient handling of data structures and computational steps to manage the dynamics of multiple interacting species effectively.

\medskip

The key differences with respect to single-species hybrid methods can be summarised as follows \cite{spill2015hybrid}:
\begin{itemize}
    \item \textbf{Multiple Species Initialisation:} Unlike single-species models, multiple species are initialized, each with its own set of parameters (diffusion coefficients, reaction rates, thresholds).
    \item \textbf{Species-Specific Domains:} Each species may have its own stochastic and deterministic domains, requiring separate management of interfaces for each species.
    \item \textbf{Multiple Interfaces:} Separate and independently moving interfaces for each species, as opposed to a single interface in the single-species model.
    \item \textbf{Coupled Reactions:} Handling reactions that involve species in different domains, ensuring accurate coupling between stochastic and deterministic regions.
    \item \textbf{Computational Complexity:} Increased complexity due to multiple sets of dynamics and interactions, requiring more efficient handling of data structures and computational steps.
\end{itemize}

With all these considerations in mind, we can now give a generic description of a multi-species hybrid algorithm:
\begin{enumerate}

    \item {Initialize Model}
\begin{itemize}
    \item \textbf{Define Spatial Domain}: Set up the spatial domain and discretize it into compartments.
    \item \textbf{Initialize Species and Parameters}: Define multiple species, their initial populations, diffusion coefficients, reaction rates, and thresholds for switching between stochastic and deterministic descriptions.
    \item \textbf{Set Up Interfaces}: Determine initial positions for interfaces between stochastic and deterministic domains for each species.
\end{itemize}

\item{Main Loop}
\begin{itemize}
    \item \textbf{Step 1: Generate Time to Next Stochastic Event}
    \begin{itemize}
        \item Use Gillespie’s algorithm to determine the time step for the next event.
    \end{itemize}
    \item \textbf{Step 2: Determine Which Stochastic Event Occurs}
    \begin{itemize}
        \item Identify the specific stochastic event that occurs based on the random number generated.
    \end{itemize}
\end{itemize}

\item{Update Deterministic Domain}
\begin{itemize}
    \item \textbf{Finite Difference Scheme}: Apply the finite difference scheme to update the deterministic regions for each species.
    \item For each species in the deterministic domain, update their concentrations using the PDEs.
\end{itemize}

\item{Calculate Interface Conditions}
\begin{itemize}
    \item \textbf{Stochastic to Deterministic Fluxes}: Calculate fluxes between stochastic and deterministic compartments for each species at the interfaces.
    \item \textbf{Local Reactions at Interfaces}: Perform local reactions involving species from both stochastic and deterministic domains.
\end{itemize}

\item{Move Interface if Necessary}
\begin{itemize}
    \item \textbf{Check Thresholds}: For each species, check if the particle number in a compartment crosses the threshold, and adjust the interface position if necessary.
    \item \textbf{Renormalize Particle Numbers}: Ensure the conservation of mass when adjusting the interfaces, and handle fractional particles appropriately.
\end{itemize}

\item{Update Time}
\begin{itemize}
    \item Increment the simulation time by the time step generated in Step 1.
\end{itemize}

\item{Repeat Until End Time}
\begin{itemize}
    \item Continue looping through the steps until the specified end time is reached.
\end{itemize}

\end{enumerate}

%%%%%%%%%
\subsection{A Multiple Species Example: Hybrid Lotka-Volterra}

The fine details concerning  the algorithm steps we just outlined are better described by working out in detail an example with two species, including how to account for fluxes between regions, interface placement and species interactions. The general case of more than two species, apart from combinatorial issues, does not bring in additional difficulties.

Therefore, the core of section~\ref{sec:Multiple_species} will be  constituted by the analysis of a hybrid model for a two species system, a classical example often described by the Lotka-Volterra equations, see Eqs.~\eqref{eq:LVPDE} below. This system includes a stochastic component where individuals interact and move on a lattice and a deterministic component described by partial differential equations (PDEs). By integrating both methods, we aim to capture the detailed behaviour of individual organisms while also modelling the population-level dynamics, as will be seen below.

We explore a predator-prey model comprising predators ($M$) and prey ($N$). We denote by $M(k)$ and $N(k)$ the number of predators and prey, respectively, in voxel $k$. We shall write $M(t,k), N(t,k)$ when we need to stress the temporal dependency.

The model is described in terms of its transition rates $W(B|A)$, which, to first order in $\Delta t$, encode the probability of a system being in state $A$ at time $t$ to switch to state $B$ during the time interval $[t,t+\Delta t)$, divided by $\Delta t$. For one-dimensional spatial systems, the state of each species is described via the state vector:
\[
\mathcal{N}= (N(k_{1}),\ldots,N(k_{max}))
\]
where $\{k_{1},\ldots,k_{max}\}$ is the indexing set of the spatial lattice. Time dependency might be explicitly included if needed. The overall state of the system is described by the joint state vector of all the species. 
%In the sequel,
Hereafter we follow the convention that, when specifying transition rates, only the components of the state vector that change during the transition are explicitly written.

The stochastic model allows each species to migrate to neighbouring lattice sites. Prey reproduce at a rate $a$, while predators have a mortality rate $c$ and reproduce by consuming prey at a rate $b$. The transition rates for these interactions are:

\begin{align}
\label{eq:transitionRatesLV}
W\left(N(k)-1,N(k\pm 1)+1|N(k),N({k\pm 1})\right) &= \frac{D_N}{h^2}N(k),\nonumber
\\
W\left(M(k)-1,M({k\pm 1})+1|M(k),M(k\pm 1)\right) &= \frac{D_M}{h^2}M(k),\nonumber
\\
W\left(N(k)+1|N(k)\right) &= a N(k),
\\
W\left(N(k)-1,M(k)+1|N(k),M(k)\right) &= b N(k) M(k),\nonumber
\\
W\left(M(k)-1|M(k)\right) &= c M(k),\nonumber
\end{align}

where $D_M$ and $D_N$ denote the diffusion coefficients of $M(k)$ and $N(k)$ respectively and $h$ is the spatial discretization size. 
For simplicity, we assume that each time a predator consumes a prey, a new predator is born. The mean-field and continuum limit lead to the classical spatial Lotka-Volterra equations:

\begin{align}\label{eq:LVPDE}
\frac{\partial n}{\partial t} &= D_N\frac{\partial^2 n}{\partial x^2} + a n - b n m,\nonumber
\\
\frac{\partial m}{\partial t} &= D_M\frac{\partial^2 m}{\partial x^2} + b n m - c m.
\end{align}

Here, $n=n(t,x)$ and $m=m(t,x)$ represent the prey and predator densities, respectively, analogous to $N(t,k)$ and $M(t,k)$ in the discrete model.

We solve Eqs.~\eqref{eq:LVPDE} using a finite difference approximation for spatial discretisation, using the same lattice as the stochastic model. For time integration, the fourth-order Runge-Kutta explicit method is used. All plots show normalised values, indicating the number of predators or prey in a box rather than their densities.

The model consists of four subdomains, determined by whether each species evolves deterministically or stochastically. For simplicity, we denote $N(k)= h n(k)$ and $M(k)= h m(k)$ where appropriate. We now specify which transition rates define the stochastic model and which PDEs correspond to the mean-field model solved in each subdomain. The interfaces between the subdomains are as described in section~\ref{sec:hybrid_algorithm}:

\begin{enumerate}

\item Deterministic Predator and Prey System. 

In this region, Eqs.~\eqref{eq:LVPDE} are solved, with the transition rates from Eqs.~\eqref{eq:transitionRatesLV} set to zero, ensuring no stochastic reactions occur.

\item Deterministic Predator and Stochastic Prey. 

The deterministic equations are:

\begin{align}\label{eq:LVPDEDetPredStochPrey}
\frac{\partial n}{\partial t} = 0,\quad \frac{\partial m}{\partial t} = D_M\frac{\partial^2 m}{\partial x^2}  - c m,
\end{align}

and the transition rates are:
\begin{align}
\label{eq:LVstochDetPredStochPrey}
W\left(N(k)-1,N({k\pm 1})+1|N(k),N({k\pm 1})\right) &= \frac{D_N}{h^2}N(k),\nonumber
\\
W\left(M(k)-1,M({k\pm 1})+1|M(k),M(k\pm 1)\right) &= 0,\nonumber
\\
W\left(N(k)+1|N(k)\right) &= a N(k),
\\
W\left(N(k)-1,M(k)+1|N(k),M(k)\right) &= b N(k) M(k),\nonumber
\\
W\left(M(k)-1|M(k)\right) &= 0.\nonumber
\end{align}

\item Stochastic Predator and Deterministic Prey. 
The deterministic equations are:

\begin{align}
\frac{\partial n}{\partial t} = D_N\frac{\partial^2 n}{\partial x^2} + a n ,\quad \frac{\partial m}{\partial t} = 0,
\end{align}

and the transition rates are:
\begin{align}
\nonumber
W\left(N(k)-1,N(k\pm 1)+1|N(k),N({k\pm 1})\right) &= 0,\nonumber
\\
W\left(M(k)-1,M({k\pm 1})+1|M(k),M(k\pm 1)\right) &= \frac{D_M}{h^2}M(k),\nonumber
\\
W\left(N(k)+1|N(k)\right) &= 0,
\\
W\left(N(k)-1,M(k)+1|N(k),M(k)\right) &= b N(k) M(k),\nonumber
\\
W\left(M(k)-1|M(k)\right) &= c M(k).\nonumber
\end{align}

\item Stochastic Predator and Stochastic Prey. 

In this scenario, both species evolve stochastically according to the transition rates defined by Eqs.~\eqref{eq:transitionRatesLV}.\\
\end{enumerate}

We now explore two scenarios within the spatial Lotka-Volterra system and compare the hybrid model to purely stochastic and deterministic models. Both scenarios involve solutions of the PDE that oscillate in space and time, but in one case, the oscillations bring the prey population so close to zero that extinction is possible in the stochastic model.

%%%%%%%%%%%%%%%%%%%%%%%%%

\subsection{Oscillatory Behaviour without Observable Extinction}
In a domain of length $L=20$ divided into $k_{max}=101$ boxes ($h=0.2$), the initial condition is a spatially homogeneous distribution of prey and predators, with $N(k,t=0)=50$ and $M(k,t=0)=5$. With Neumann boundary conditions, Eqs.~\eqref{eq:LVPDE} maintain spatial homogeneity at all times, and both populations oscillate in time.

\begin{figure}[h!]
\centering
\includegraphics[width=0.60\linewidth]{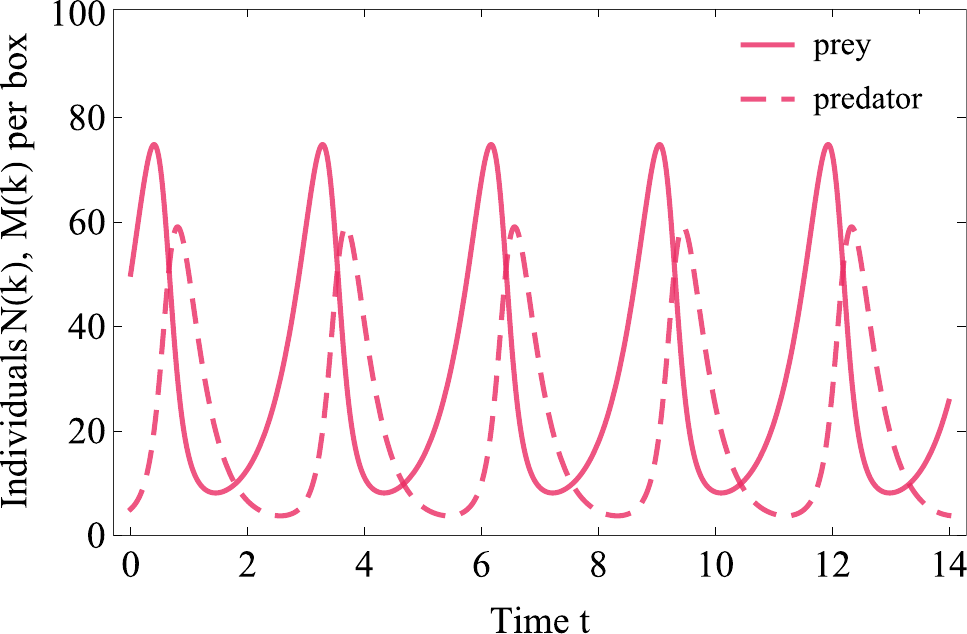}
\caption{The solution of the Lotka-Volterra Eq.~\eqref{eq:LVPDE} with parameters $D_N=D_M=1$, $a=2$,  $b=0.1$, $c=3$ for spatially homogeneous initial conditions $N(t=0,k)=50$, $M(t=0,k)=5$. The plot shows the time evolution of the number of prey and predators in any given box $k$ in the discretisation, for better comparison with the stochastic model. Since the diffusion terms do not contribute in the spatially homogeneous case, this solution is identical to the solution of the Lotka-Volterra ODEs.} \label{fig:flatOscillation_LVPDE}
\end{figure}

Figure~\ref{fig:flatOscillation_LVPDE} illustrates typical results, showing that peaks in prey numbers are followed by peaks in predator numbers. For the chosen parameter values, the minimum number of individuals of either species remains sufficiently large to make extinction in the stochastic model unlikely.

\begin{figure}[h]
\subfloat[Spatial Profile Stochastic Model]
{
\label{fig:flatOscillationStoch_1}
\includegraphics[width=0.48\linewidth]{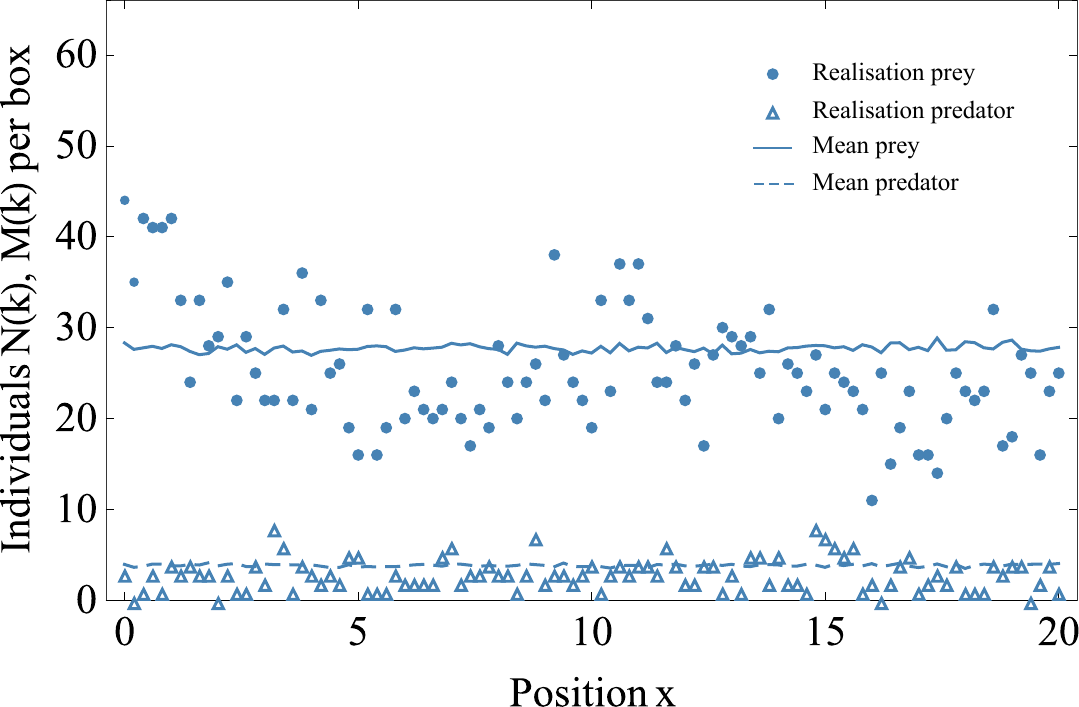}
}
\subfloat[Time Evolution Stochastic Model]
{
\label{fig:flatOscillationStoch_2}
\includegraphics[width=0.48\linewidth]{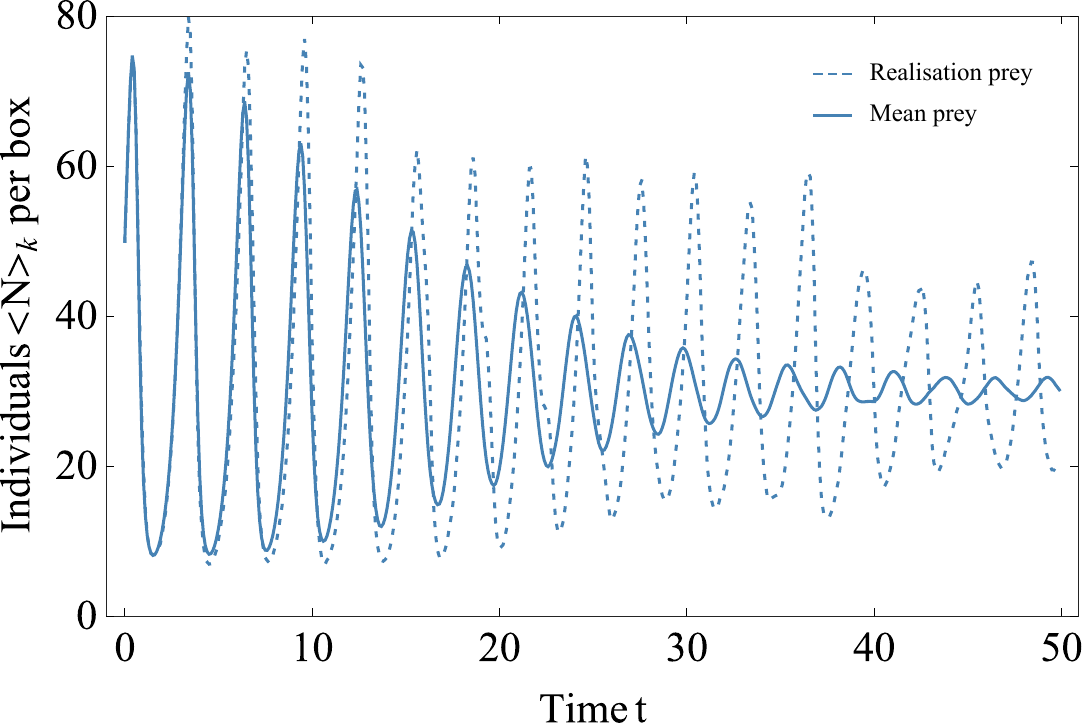}
}\\
\subfloat[Spatial Profile Hybrid $\Theta=10$]
{
\label{fig:flatOscillationHybrid10_1}
\includegraphics[width=0.48\linewidth]{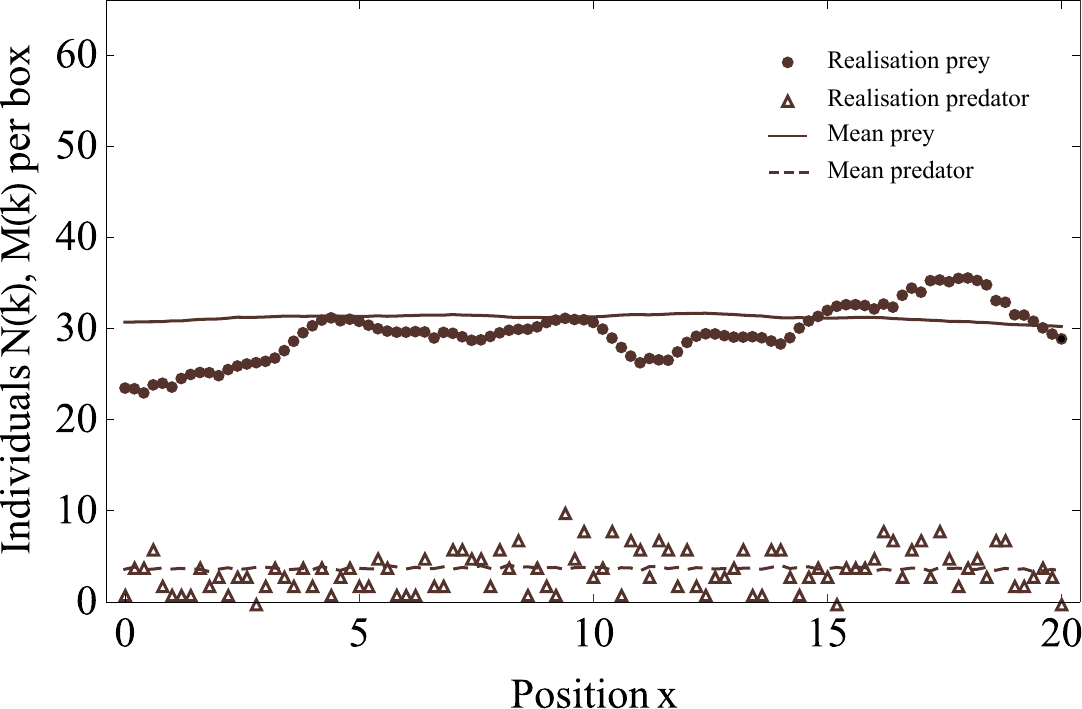}
}
\subfloat[Time Evolution Hybrid $\Theta=10$]
{
\label{fig:flatOscillationHybrid10_2}
 \includegraphics[width=0.48\linewidth]{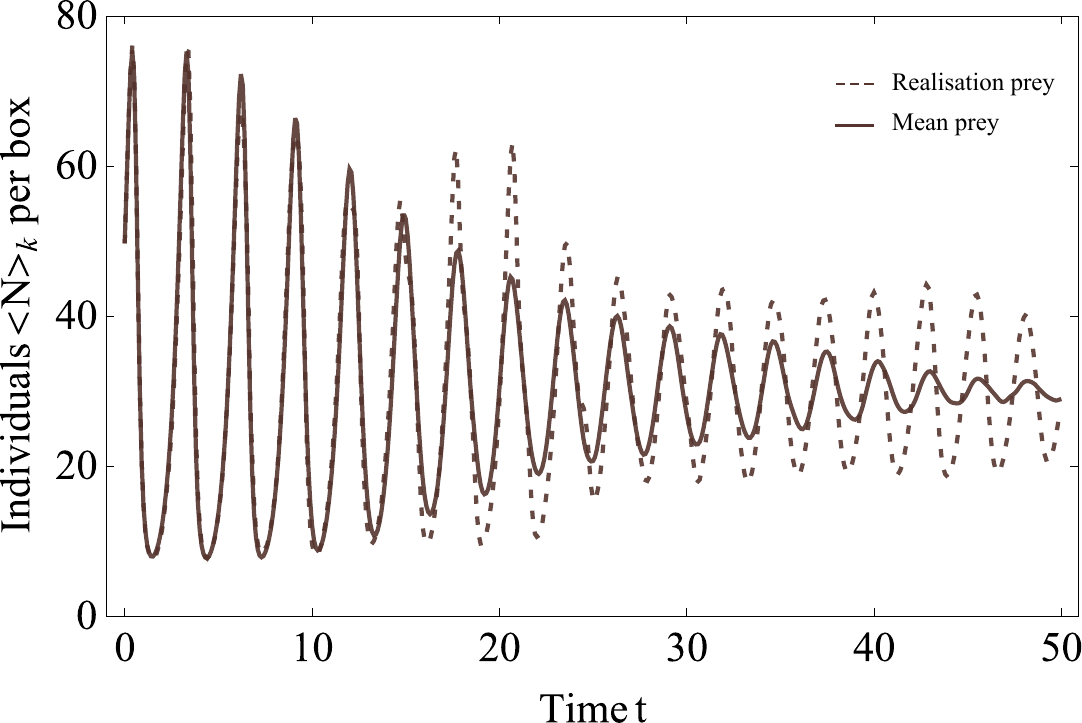}  

}

\caption{Simulations of the spatial Lotka-Volterra Model with parameters $D_N=D_M=1$, $a=2,b=0.1,c=3$, $k_{max}=101, h=0.2$ and initial values $N(t=0,k)=50, M(t=0,k)=5$ for $k=1,\dots,k_{max}$. We compare \protect\subref{fig:flatOscillationStoch_1}-\protect\subref{fig:flatOscillationStoch_2}, the stochastic model, to \protect\subref{fig:flatOscillationHybrid10_1}-\protect\subref{fig:flatOscillationHybrid10_2}, the hybrid model with thresholds $\Theta=10$. The figures on the left show the spatial profile at time $t=4.1$ of the number of predators and prey of a single realisation as well as the mean of $256$ different realisations, whereas the figures on the right show the time evolution of the spatial average of numbers of prey in a single realisation and the mean of realisations. The corresponding PDE solution is shown in Figure~\ref{fig:flatOscillation_LVPDE}.}\label{fig:flatOscillation}
\end{figure}

The results for the stochastic and hybrid models, using two different threshold values, are discussed in \cite{spill2015hybrid} observing differences in the result if the threshold is large. Figure~\ref{fig:flatOscillation} compares purely stochastic and hybrid methods with $\Theta=10$. The left column illustrates the spatial distribution of predator and prey populations within a specific box at time $t=4.1$, both for an individual simulation and the average of 256 simulations. When comparing the predator and prey populations, it is evident that the average values for both the stochastic and hybrid models are quite similar for the given threshold (refer to Figures~\ref{fig:flatOscillation} (a) and (c)). The individual simulations, however, show significant differences. As expected, the stochastic model exhibits considerable noise in predator and prey populations across the domain, whereas the hybrid model shows reduced noise once population sizes exceed the threshold. Figure~\ref{fig:flatOscillation} (c) demonstrates that predator numbers are above the threshold, resulting in a smooth spatial distribution, although this distribution is not homogeneous, unlike the average profile.

The right column of Figure~\ref{fig:flatOscillation} presents the temporal evolution of the spatial average number of prey, calculated as $\left\langle N\right\rangle_k=\frac{1}{k_{max}}\sum_{k=1}^{k_{max}}N(k)$. The mean values for the stochastic model (Figure~\ref{fig:flatOscillation}~\subref{fig:flatOscillationStoch_2}) closely resemble those for the hybrid model, (Figure~\ref{fig:flatOscillation}~\subref{fig:flatOscillationHybrid10_2}). In all cases, the system exhibits oscillatory behaviour around $n=30$, the steady state of the corresponding PDE, with a diminishing amplitude over time. This contrasts with the PDE dynamics (Figure~\ref{fig:flatOscillation_LVPDE}), where oscillation amplitudes remain constant. Stochastic fluctuations cause the oscillations to desynchronize and average out, leading to a more pronounced damping effect in the mean of all realisations compared to individual ones, where amplitudes may vary over time. This indicates that the PDE does not accurately represent the mean over different realisations. Additionally, the spatial mean predator plots for the stochastic and hybrid models with different thresholds may look different; however, this variability is also seen in different realisations of the same model. Therefore, quantitative measures are necessary for a proper comparison (see Table \ref{tab:LVoscillations_statisticalMeasures}).

\begin{table}[h]
\centering
\begin{tabular}{|c|c|c|c|c|}\hline
 &  Stochastic & Hybrid $\Theta=10$ &Mean Field \\ \hline
$\mean{\left\langle {N}\right\rangle}_{t,k}$ &	$30.50\pm 0.05$ & $30.30\pm 0.16$ &$30.47$\\
$\mean{\left\langle {M}\right\rangle}_{t,k}$&	$20.35\pm 0.04$  & $20.22\pm 0.13$ &$20.20$\\
$\sqrt{\mean{\left\langle {N}\right\rangle^2_{t,k}}-\mean{\left\langle {N}\right\rangle}_{t,k}^2}$ &	$16.0 \pm 0.3$ & $15.4\pm 0.4$ &\\
$\sqrt{\mean{\left\langle {M}\right\rangle^2_{t,k}}-\mean{\left\langle {M}\right\rangle}_{t,k}^2}$ &	$12.8 \pm 0.3$  & $12.3\pm 0.4$ &\\\hline
 \end{tabular}
  \caption{Average prey population over space, time, and 256 realisations, along with predator numbers. The standard deviation, for prey and predators, are also displayed. Results for the corresponding mean field model, where there are no different realisations, and thus no standard deviation, are included.}
 \label{tab:LVoscillations_statisticalMeasures}
 \end{table}

We calculate the spatial and temporal average of the mean number of prey across 256 different realisations,

\[
\mean{\left\langle N \right\rangle}_{t,k} = \frac{1}{256 k_{max} t_{max}} \sum_{r=1}^{256} \sum_{k=1}^{k_{max}} \int_0^{t_{max}} N_r(t,k) dt, \label{mean_average}
\]
over a simulation time of $t_{max}=50$, and similarly for predators. This provides a single value with high statistical significance, facilitating easier comparison as individual time points may have oscillations that are out of synchrony. The standard deviation of the spatio-temporal averages, $\sqrt{\mean{\left\langle N \right\rangle^2_{t,k}} - \mean{\left\langle N \right\rangle}_{t,k}^2}$, for prey and predators shows close agreement between the stochastic model and the hybrid model with  $\Theta=10$. The hybrid model effectively reproduces the stochastic measures, such as standard deviation, with better accuracy for higher threshold values.

\subsection{Extinction and Blow-up} \label{sec:extinctionBlowup}

\begin{figure}[h!]
\subfloat{
\includegraphics[width=0.48\linewidth]{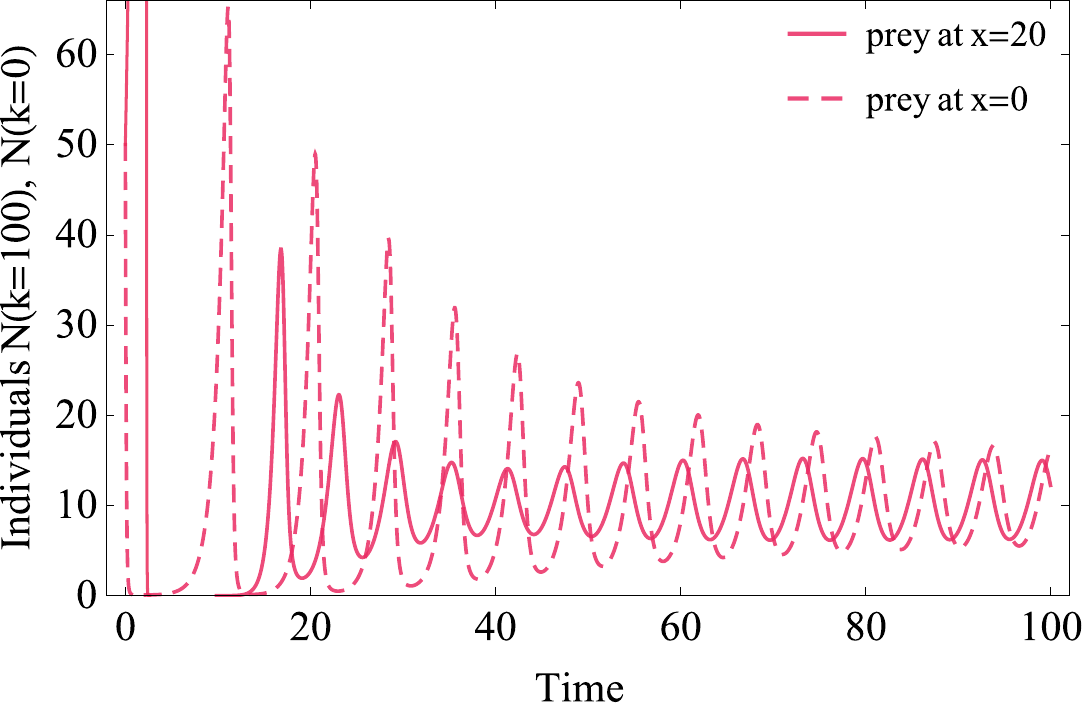}
}
\subfloat{
\includegraphics[width=0.48\linewidth]{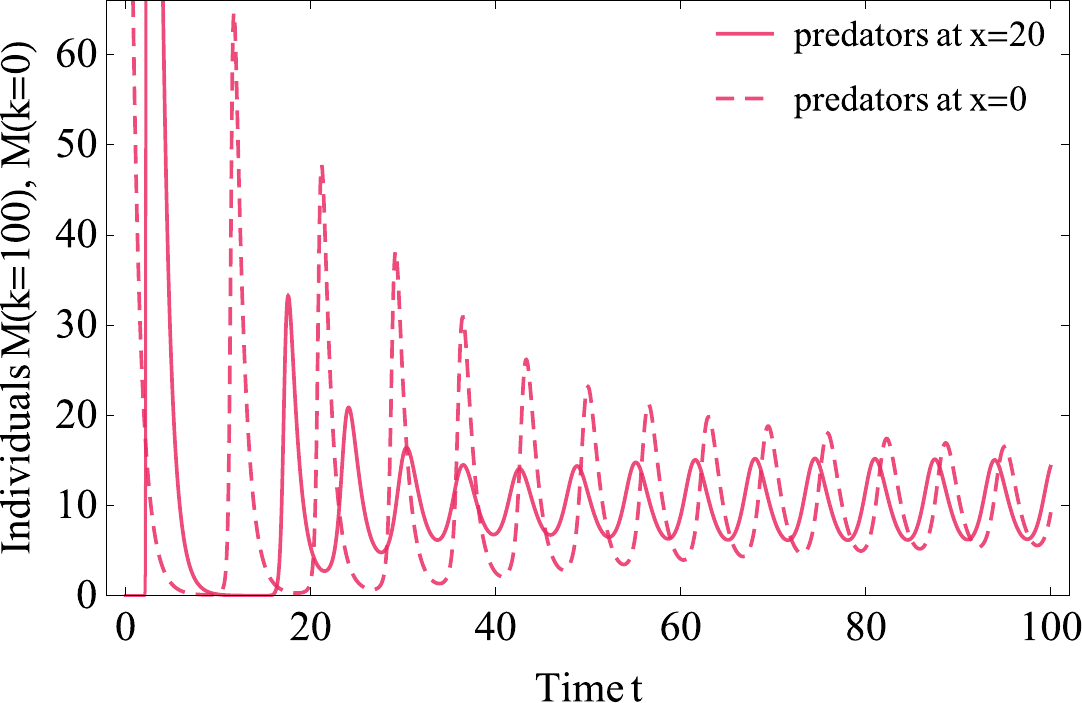}
}
\caption{Number of prey (left) and predators (right) at the boundaries ($x=0$ and $x=20$) over time, obtained from the deterministic model Eq.~\eqref{eq:LVPDE}. The initial sharp rise in prey numbers at $x=20$ is due to the lack of predators in that region and is followed by a predator peak. Early oscillations show population numbers close to zero, indicating possible extinction in the stochastic model. Later times show oscillations with minima sufficiently above zero, reducing the likelihood of extinction.}\label{fig:extinction_PDE_two_locations}
\end{figure}

In this analysis, we set $k_{max}=101$ and we consider an initial condition where prey are uniformly distributed, $N(t=0,k)=50$ for $k=1,\dots,k_{max}$, while predators are concentrated on the left side of the domain, with $M(k)=100$ for $k=1,\dots,9$, $M(10)=98$, $M(11)=50$, $M(12)=2$, and $M(k)=0$ for $k=13,\dots,k_{max}$. The boundary conditions are of Neumann type, and the parameters are set to $D_N=D_M=1$, $a=1$, $b=0.1$, $c=2$, with a domain length $L=20$. This setup models a predator invasion into a prey population, causing spatial and temporal fluctuations even in a deterministic framework (Figure \ref{fig:extinction_PDE_two_locations}).

Figure \ref{fig:extinction_PDE_two_locations} illustrates the variations in predator and prey numbers at the boundaries ($x=0$, $x=20$) over time for the mean-field model. During the initial oscillations, both predator and prey populations dip close to zero. In the stochastic model, where population numbers are integers, values below one in the deterministic model suggest a high likelihood of extinction. Later on, regular oscillations with minima significantly above zero are observed, indicating that if extinction occurs, it is most probable during early times.

\begin{figure}[h!]
\subfloat[Stochastic Model, Oscillation]
{
\label{fig:extinction_stoch_mean_osc}
\includegraphics[width=0.48\linewidth]{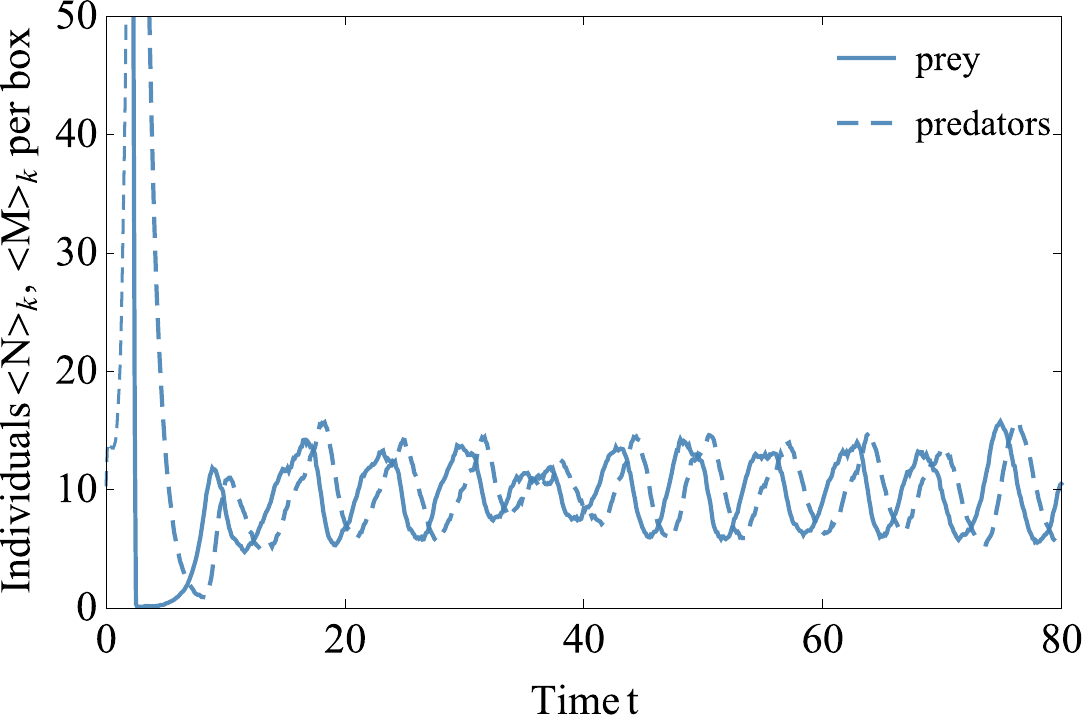}
}
\subfloat[Hybrid Model, Oscillation]
{
\label{fig:extinction_hybrid_10_mean_osc}
\includegraphics[width=0.48\linewidth]{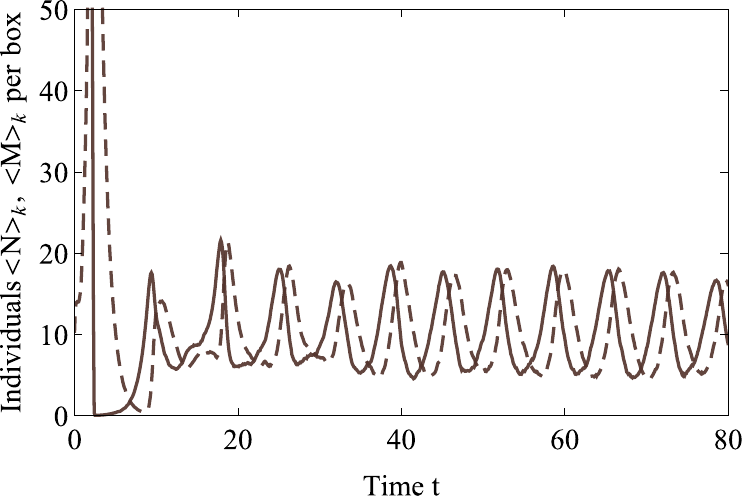}
}\\
\subfloat[Stochastic Model, Extinction]
{
\label{fig:extinction_stoch_mean_ext}
\includegraphics[width=0.48\linewidth]{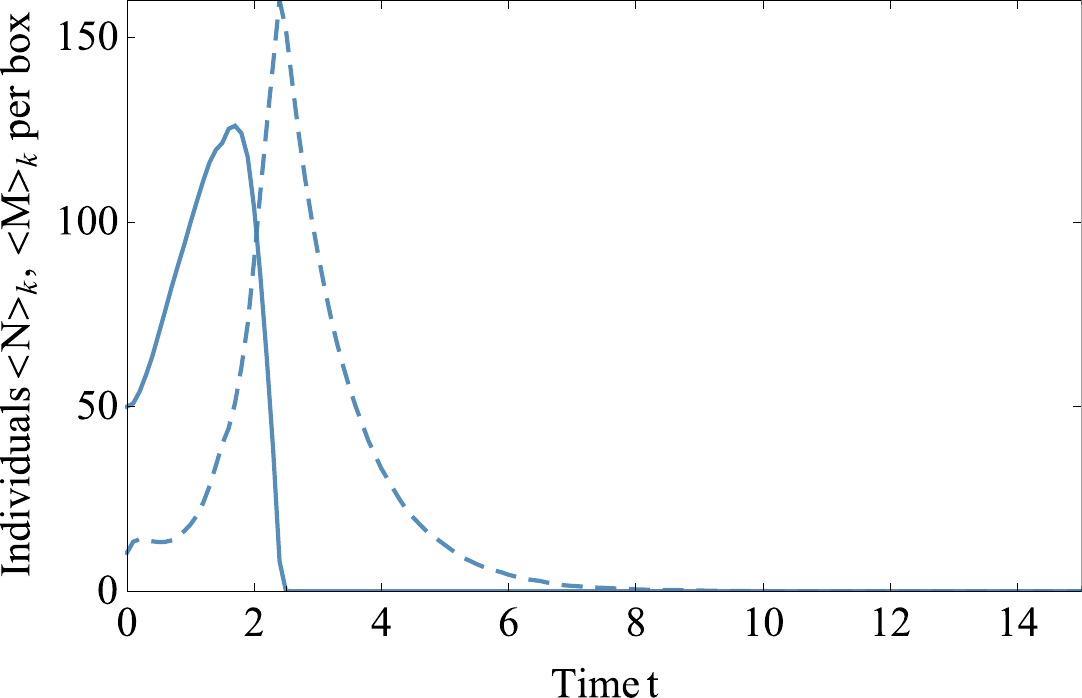}
}
\subfloat[Hybrid Model, Extinction]
{
\label{fig:extinction_hybrid_10_mean_ext}
 \includegraphics[width=0.48\linewidth]{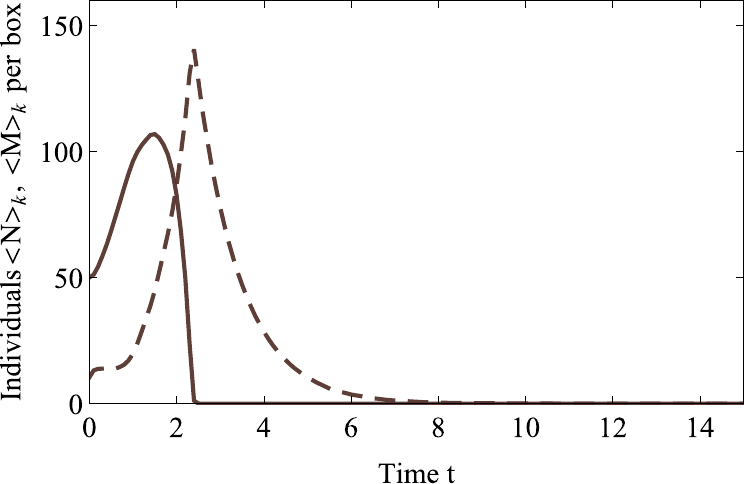}  

}\\
\subfloat[Stochastic Model, Blow-Up]
{
\label{fig:extinction_stoch_mean_blow}
\includegraphics[width=0.48\linewidth]{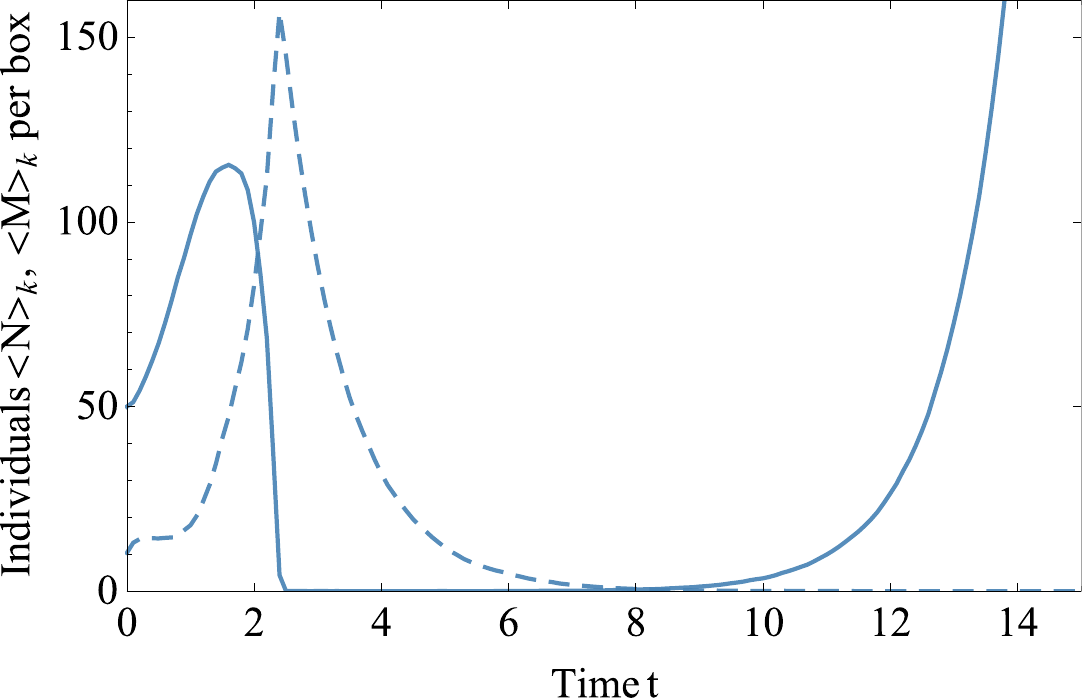}
}
\subfloat[Hybrid Model, Blow-Up]
{
\label{fig:extinction_hybrid_10_mean_blow} 
\includegraphics[width=0.48\linewidth]{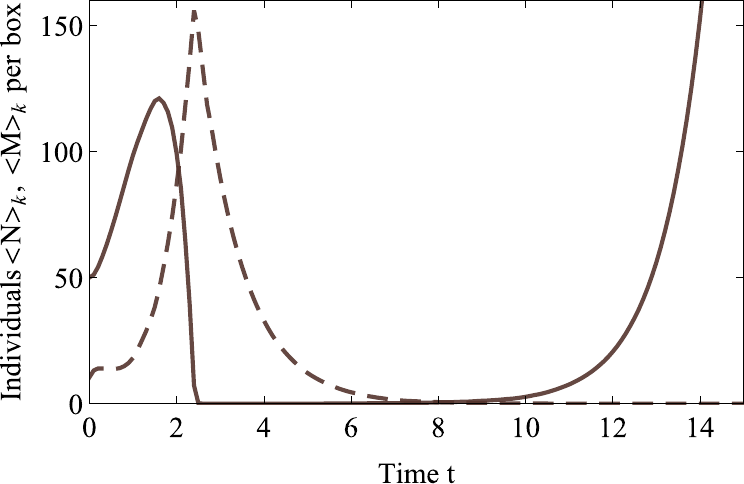}  
}
\caption{Simulations of the spatial Lotka-Volterra model with parameters $D_N=D_M=1$, $a=1$, $b=0.1$, $c=1$, $L=20$, and $k_{max}=101$. The spatial average $\left\langle N\right\rangle_k=\frac{1}{k_{max}}\sum_{k=1}^{k_{max}}N(k)$ of the number of predators and prey in a box is plotted over time for three realisations of the stochastic model \protect\subref{fig:extinction_stoch_mean_osc},\protect\subref{fig:extinction_stoch_mean_ext},\protect\subref{fig:extinction_stoch_mean_blow}, and three realisations of the hybrid model, \protect\subref{fig:extinction_hybrid_10_mean_osc},\protect\subref{fig:extinction_hybrid_10_mean_ext},\protect\subref{fig:extinction_hybrid_10_mean_blow}, with a threshold of $\Theta=10$. These realisations show three different outcomes: oscillatory behaviour, extinction of both species or predator extinction followed by prey population explosion.} \label{fig:LV_extinction_hybrid_mean}
\end{figure} 

Figure~\ref{fig:LV_extinction_hybrid_mean} confirms these expectations. Figure~\ref{fig:LV_extinction_hybrid_mean}~\subref{fig:extinction_stoch_mean_osc} demonstrates that in the stochastic model, the spatial average of predator or prey numbers can exhibit oscillations similar to those in the deterministic model. However, extinction events can also occur. Figure~\ref{fig:LV_extinction_hybrid_mean}~\subref{fig:extinction_stoch_mean_ext} shows that if prey die out first, predators will necessarily follow. Conversely, if predators go extinct first, prey numbers will rapidly increase (see Figure~\ref{fig:LV_extinction_hybrid_mean}~\subref{fig:extinction_stoch_mean_blow}). Figures~\ref{fig:LV_extinction_hybrid_mean}~\subref{fig:extinction_hybrid_10_mean_osc}, \subref{fig:extinction_hybrid_10_mean_ext} and \subref{fig:extinction_hybrid_10_mean_blow} demonstrate that the hybrid model can replicate each of these three distinct scenarios. Since the mean and standard deviation statistics are not meaningful if the prey population explodes, we compare the extinction event frequencies in the stochastic and hybrid models.

\begin{table}[h]
\centering
 \begin{tabular}{|c|c|c|c|c|c|}
 \hline
 &  Stochastic & Hybrid 100& Hybrid 50 & Hybrid 25 & Hybrid 10\\ \hline
 Prey & 0.46 & 0.45 & 0.46 & 0.45 & 0.43 \\ 
 Predator & 0.59 & 0.57 & 0.59 & 0.61 & 0.56 \\
 \hline
 \end{tabular}

\caption{Extinction probability for prey and predator populations in the Lotka-Volterra system for the described scenario. This probability is calculated by conducting 256 simulations for each model (stochastic and hybrid) with thresholds of 10, 25, 50, and 100, and recording at time $t=80$ whether a population is extinct.}
 \label{tab:extinction}
\end{table}

The results in Table \ref{tab:extinction} indicate that the hybrid model with a threshold of $\Theta=10$ or higher has a similar extinction probability as the stochastic model. Therefore, the hybrid model is a good approximation of the stochastic model in this case. Finally, we evaluate the performance gain achieved by using the hybrid model.

\section{Hybrid Method for Convergent Reaction-Diffusion Master Equation} \label{sec:hybrid_cRDME}

Stochastic reaction-diffusion systems are commonly modelled using the compartment-based Reaction-Diffusion Master Equation (RDME). This method divides the domain into non-overlapping compartments (voxels) to capture spatial inhomogeneities. In this mesoscopic description, each molecule's position is known up to the voxel size, with molecules within the same voxel treated as indistinguishable and well-mixed. Diffusion between compartments is modelled as a continuous-time random walk, and chemical reactions occur independently within each compartment. The RDME approach is based on the assumption of the existence of propensity functions (which is an alternative term to refer to transition rates in the context of chemical kinetics). For well-mixed systems, propensities are functions of the number of reactant molecules present in the system. However, when molecular species in compartments are present in low quantities, the well-mixed assumption ceases to be valid, and a straightforward generalisation of the Master Equation to reaction-diffusion systems may produce incorrect results. Specifically, there exists a lower bound for the size of each compartment below which the probability of bimolecular reactions vanishes \cite{gillespie2014validity,smith2016breakdown}. This results in waiting times for bimolecular reactions tending to infinity as the lattice size decreases, causing inaccuracies with finer meshes. Therefore, for the RDME approach to be valid, the mesh size must be significantly larger than the reaction radius between molecules.

Two strategies have been employed to address this issue. One method introduces generalised reaction rates (propensities) for bimolecular reversible reactions of the type $A + B \rightleftarrows C$, which are allowed to occur between neighbouring voxels \cite{hellander2016reaction, erban2009stochastic}. To distinguish between reactions occurring within the same voxel and those between neighbouring ones, an additional correcting parameter is introduced \cite{hellander2016reaction}. This parameter can be derived by matching the mean binding-rebinding times at micro- and mesoscales, resulting in the corresponding reaction rates for the Smoluchowski interaction model in 2D and 3D. This modified RDME is accurate down to a smaller lower mesh size bound, approximately equal to the interaction radius.

A different and more general approach was proposed by \cite{isaacson2013convergent}. This work introduces the convergent RDME (cRDME) for bimolecular reactions of type $A + B \rightarrow C$ within the Doi interaction model. The cRDME formulation allows interaction between molecules from different voxels that lie within a predefined reaction radius. This addresses the issue of vanishing reaction rates as voxel size approaches zero. The corresponding reaction rates are obtained from a finite volume discretisation of the spatially continuous stochastic reaction-diffusion model. It was demonstrated that for large mesh sizes, the cRDME coincides with the classical RDME, thus extending the existing approach. The cRDME can be simulated using the traditional Stochastic Simulation Algorithm (SSA) \cite{gillespie1976general} or its more efficient formulations, such as the Next Subvolume (NSV) method \cite{elf2004spontaneous}.

While mesoscopic approaches offer computational advantages over microscopic molecular-based simulations, they often encounter practical limitations in real-world applications. As a result, alternative methods are frequently needed to simulate complex reaction-diffusion systems more effectively. In this section, we introduce one such approach, which consists in formulating a hybrid method for the cRDME. As an application, we choose the stochastic Fisher-Kolmogorov system, which is particularly well-suited to study the implementation of hybrid methods, as it exhibits travelling wave behaviour \cite{moro2004,spill2015hybrid,yates2015pseudo,delacruz2017coarse}. We begin with a brief description of this system within the convergent framework in section~\ref{subsec:Fisher_cRDME}. In section~\ref{subsec:Hybrid_cRDME}, we proceed with a description of our implementation of a hybrid method for cRDME. We finally compare its performance with the purely stochastic simulations via the NSV method in section~\ref{subsec:Results_cRDME}.

\subsection{Convergent Formulation of the Stochastic Fisher-Kolmogorov Model}\label{subsec:Fisher_cRDME}

The stochastic Fisher Kolmogorov (sFK) system for a single species is described by two processes, corresponding to branching (birth) and binary annihilation 
\begin{align*}
A &\xrightarrow{\text{$\bar \lambda$}} 2A, \\
2A &\xrightarrow{\text{$\bar \mu$}}  \emptyset,
\end{align*}
\noindent where $\bar \lambda$ and $\bar \mu$ characterise mesoscopic birth and annihilation (death) rates, respectively. They can be obtained from the macroscopic reaction rates, $\lambda$ and $\mu$, respectively, by appropriate scalings, which we describe in more detail below. We denote by $\mathbf{N}(t)$ the total number of molecules present in the system. Starting from a positive initial condition, $\mathbf{N}_0=\mathbf{N}(t=0)$, this system equilibrates when the population reaches its carrying capacity, given by the ratio of the macroscopic reaction rates, $\mathcal{K}=\lambda/ \mu$. If the spatial distribution of molecules plays a relevant role, this description must be complemented by diffusion dynamics.

The birth reaction, being first-order (unimolecular), does not present convergence issues. However, second-order annihilation reactions require special consideration. Whether using the RDME or its convergent version (cRDME), the mesoscopic rate for bimolecular annihilation, $\bar{\mu}$, is obtained by scaling the corresponding macroscopic rate with the lattice spacing and the system size, $\Omega$ \cite{spill2015hybrid}. The system size parameter has units of area (volume) and allows to perform conversions between the number of molecules and molecular concentrations, i.e. $n = \mathbf{N}/{\Omega}$. The implementation of the convergent formulation also requires that the molecules in the sFK system interact between different voxels within the reaction radius \cite{isaacson2013convergent}. Diffusion of molecules will be modelled in the standard way as a continuous-time unbiased random walk between neighbouring compartments. We will consider reflecting zero-flux boundary conditions to represent a closed system. For simplicity, we will present our results in one dimension. Nonetheless, it is straightforward to extend the formulation of the convergent sFK system and the hybrid method described later to higher dimensions.

Consider a one-dimensional domain $\mathcal{D}=[-l,l]$. We partition $\mathcal{D}$ into  $2K$ equal voxels or compartments of size $h=l/K$. Then a voxel $k$ is given by an interval $v_k=[(k-1)h,\ kh]$ for  $k=\{-K+1,\ldots,K-1,K\}$.
 We denote by $N(k)$ the number of particles in voxel $k$. Molecules in each voxel are assumed to be well-mixed and identical. The vector $\mathcal{N}=(N(-K+1) \ldots N(K)) $ describes the current state of the system. Its dynamics are then described by a Master Equation of the form:
\begin{align} \label{RDME_General}
\frac{d P(t,\mathcal{N})}{d t}=\displaystyle \sum_{\widetilde{\mathcal{N}}} \af{W (\mathcal{N}|\widetilde{\mathcal{N}}) P(t,\widetilde{\mathcal{N}}) - W(\widetilde{\mathcal{N}}|\mathcal{N}) P(t,\mathcal{N})}.
\end{align}
\noindent Here, $P(t,\mathcal{N})$ represents the probability distribution of system states at time $t$. The transition rates, $W$, must be properly specified according to the cRDME approach introduced in \cite{isaacson2013convergent} to ensure its convergence. 

\subsubsection{Mesoscopic Reaction Rates}\label{subsubsec:MesoRates}

\textbf{Diffusion ~~~~} We describe the random movement of molecules between voxels in the usual manner using the following rates:
\begin{align} \label{Wdiff}
W_d \of{N(k)-1,N(k\pm 1) + 1 | N(k),N(k\pm 1)}=\overline{D}N(k).
\end{align}
\noindent Since these rates are linear with respect to $N(k)$, they are not affected by convergence issues and therefore require no modification. The scaled diffusion coefficient, $\overline{D}$, equals $D h^{-2}$.

\smallskip
\noindent
\textbf{Birth ~~~~} We assume that birth events occur at a constant rate per molecule. The associated propensity is given as follows
\begin{align} \label{Wbirth}
W_b \of{N(k)+1 | N(k)}=\overline{\lambda} N(k),
\end{align}
\noindent with $\overline{\lambda}=\lambda$. When a birth event occurs, the new molecule is placed in the same voxel $k$.  

\smallskip
\noindent
\textbf{Annihilation ~~~~} In order to account for non-linear (i.e. multi-molecular) reactions, we use the Doi model of partial absorption. The Doi model for molecular interaction assumes that annihilation can only occur if the distance between molecules is less than the prescribed reaction radius, $R_a$. Since molecule positions are known only up to the compartment size, particles from voxels $k$ and $m$ annihilate each other at a reaction rate determined by three factors: the rescaled death rate, $\mu$, a geometric correction factor, $p(k,m)$, and a combinatorial factor associated with second-order kinetics \cite{gillespie1976general}.

The rescaled annihilation rate, $\overline{\mu}$, is given by $\overline{\mu}=\mu/\of{2 h \Omega}$ for $m \neq k$. Here, $\mu$ is the macroscopic annihilation rate and $\Omega$ is the system size. The factor of $1/2$ appears due to the fact that annihilation reactions are symmetric. Thus, for a pair of voxels $k$ and $m$, the total rate is equal to $2 \overline{\mu}$, where this factor comes into play to cancel the additional $1/2$. For annihilation reactions within the same voxel, i.e. $m=k$,  $\overline{\mu}=\mu/\of{h \Omega}$ \cite{spill2015hybrid}.

The correction factor, $p(k,m)$, accounts for the statistical effects of partial coverage of compartments by the reaction region. Binary reactions occur if and only if $v_{m} \cap B_{R_a}(k) \neq \emptyset$, where $B_{R_a}(k)$ is a ball of radius $R_a$ centred at the midpoint of voxel $k$. Thus, the so-called \textit{reaction ratio}, $p(k,m) = |v_{m} \cap B_{R_a}(k)| / |v_{m}|$, represents the probability of the voxels $k$ and $m$ interacting when they are only partially covered by the reaction region. Here, $| \cdot |$ denotes the length of the one-dimensional interval. In the 1D case, the explicit formula for $p(k,m)$ reads as follows:
\begin{align} \label{eq:pkm}
p(k,m) =
\begin{cases*}
	\min \of{ \frac{2 R_a}{h},1}, & \text{if } m=k; \\
	\max\af{ 0, \min \of{ \frac{R_a}{h} -\mid m-k\mid + \skp 0.5,1}}, &\text{otherwise}.\\
\end{cases*}
\end{align}
\noindent Similarly, the probability of selecting two voxels must be appropriately defined. Although there is no unique way to define this probability, any reasonable choice should result in similar dynamics. For simplicity, we choose a uniform distribution. Thus, the probability for an annihilation reaction between two voxels within the reaction radius to occur is equal to $h/\of{2 R_a}$. The factor of $1/2$ appears because reactions are possible on both sides of the voxel.

Finally, according to the law of mass action, the resulting reaction rate is proportional to the number of all possible combinations of molecules that participate in the reaction. In the case of distinct voxels, it is simply the product, $N(k) N(m)$, while within a single voxel, the number of distinct combinations is $1/2 N(k) (N(k) - 1)$. We note the factor of $1/2$, which was previously absent in the rescaled reaction rate $\bar \mu$ for $k=m$, but now appears for the number of molecule combinations. Summarising, the annihilation transition rate in the convergent formulation is given by the following expression:
\begin{align} \label{Wann}
& W_a \of{N(k)- 1,N(m)-1|N(k),N(m)} \\ \nonumber
& = \frac{\mu}{4 R_a \Omega} \skp p(k,m)
\begin{cases*}
	 N(k) N(m), & $ \text{ if }  \ m \neq k$; \\
	 N(k) (N(k) - 1), & $ \text{ if }  \ m = k$. \\
\end{cases*}
\end{align}
With the reaction rates given by Eqs.~\eqref{Wdiff}, \eqref{Wbirth} and \eqref{Wann}, the cRDME for the sFK system reads: 
\begin{align}
\begin{split}
& \frac{d P(\mathcal{N},t)}{d t} \\ &=\underbrace{\displaystyle \sum_{k, m=k \pm 1} (\mathbf{\mathcal{E}_k^1} \mathbf{\mathcal{E}_m^{-1}} -1) W_d\of{N(k)-1,N(m) + 1 | N(k),N(m)} P(\mathcal{N},t)}_{\text{diffusion}} \\ 
& + \underbrace{\displaystyle \sum_{k} (\mathbf{\mathcal{E}_k^{-1}} - 1) W_b \of{N(k)+1|N(k)} P(\mathcal{N},t)}_{\text{birth}} \\
&+ \underbrace{\displaystyle \sum_{k, m} (\mathbf{\mathcal{E}_k^{1}} \mathbf{\mathcal{E}_m^{1}} - 1) W_a \of{N(k)- 1,N(m)-1|N(k),N(m)} P(\mathcal{N},t)}_{\text{annihilation}}.
\end{split}
\label{eq:MasterEquation}
\end{align}
\noindent Here, we used the shift operator, $ \mathbf{\mathcal{E}_k^r}$, which `shifts' the number of molecules in voxel $k$ by $r$ units:
\begin{align*}
 \mathbf{\mathcal{E}_k^r} f(N(-K+1), \ldots , N(k), \ldots , N(K)) = f(N(-K+1), \ldots , N(k)+r, \ldots , N(K)).
 \end{align*}
 
Eq.~\eqref{eq:MasterEquation} with the transition rates specified above represents the cRDME for the sFK system. Its limit for $h \rightarrow 0$ is well-defined since bimolecular reactions do not vanish. Smaller lattice size simply implies molecular interactions between higher-order neighbouring compartments.

\subsubsection{Macroscopic Limit}

To derive the mean-field description for this system, we define the concentration of molecules in voxel $k$ as $n_k={N(k)}/\of{\Omega \skp h}$. In the limit $h \rightarrow 0$, this discretisation yields a continuous concentration function, $n(x,t)$.

For the standard Fisher-Kolmogorov system \cite{breuer1994fluctuation}, the mean-field limit equation is given as follows:
\begin{align*}
\frac{\partial n}{\partial t}(t,x) = D \frac{\partial^2 n }{\partial x^2}(t,x) + \lambda n(t,x) - \mu n^2(t,x).
\end{align*}
In the convergent formulation corresponding to cRDME from Eq.~\eqref{eq:MasterEquation}, the non-locality of annihilation events results in the integral term for this reaction
\begin{align} \label{eq:NLFK}
\frac{\partial n}{\partial t}(t,x) = D \frac{\partial^2 n }{\partial x^2}(t,x) + \lambda n(t,x) - \mu n(t,x) \displaystyle \int_{\mathcal{D}} b(x-y) n(t,y) dy,
\end{align}
\noindent with $b(x,y) = b(x-y)$ being a symmetric kernel called the \textit{influence function}. For the uniform probability distribution used in the annihilation transition rate, this kernel function has the following expression:
\begin{align} 
\label{eq:Kernel}
b(x - y)=\frac{1}{2 R_a} \mathbb{1}_{\mathcal{R}(x)}(y)~.
\end{align}
\noindent Here, the set $\mathcal{R}(x) = \uf{y: ~ |x - y|< R_a}$ and $\mathbb{1}_{\mathcal{R}(x)}$ denotes the indicator function for this set. 

It can be proven that the space-homogeneous stationary solution of this non-local Fisher-Kolmogorov integro-partial differential equation (IPDE) coincides with the stationary state of the standard Fisher-Kolmorogov system, given by $\lambda/\mu$.

We complement Eq.~\eqref{eq:NLFK} with no-flux boundary conditions for $x \in \partial \mathcal{D} = \uf{-l, l }$, along with an initial condition for the initial molecule distribution:
\begin{align}
\label{eq:boundary}
&\frac{\partial n}{\partial x}(t,-l)=\frac{\partial n}{\partial x}(t,l)=0~, \\
\label{eq:initial}
&n(t=0,x)=\phi(x)~.
\end{align}

\subsection{Hybrid Method for cRDME}\label{subsec:Hybrid_cRDME}

The key ingredient of a wide class of hybrid methods to simulate reaction-diffusion systems \cite{spill2015hybrid,harrison2016hybrid,moro2004} consists in defining an interface region within which the macroscopic mean-field and stochastic descriptions coexist. The coupling between these descriptions is achieved by prescribing appropriate boundary conditions at the boundaries of the interface region. Although the exact formulation of these boundary conditions varies among different hybrid methods, their aim is to minimise the effects due to the presence of interfaces. Similarly, our primary goal will be to ensure that the interface region does not impact the system's dynamics. This will be achieved by balancing the fluxes of molecules transitioning between the subdomains.

\begin{figure}[ht]
\subfloat[]
{
\label{fig:HybridDomain_A}
\includegraphics[width=0.5919\linewidth]{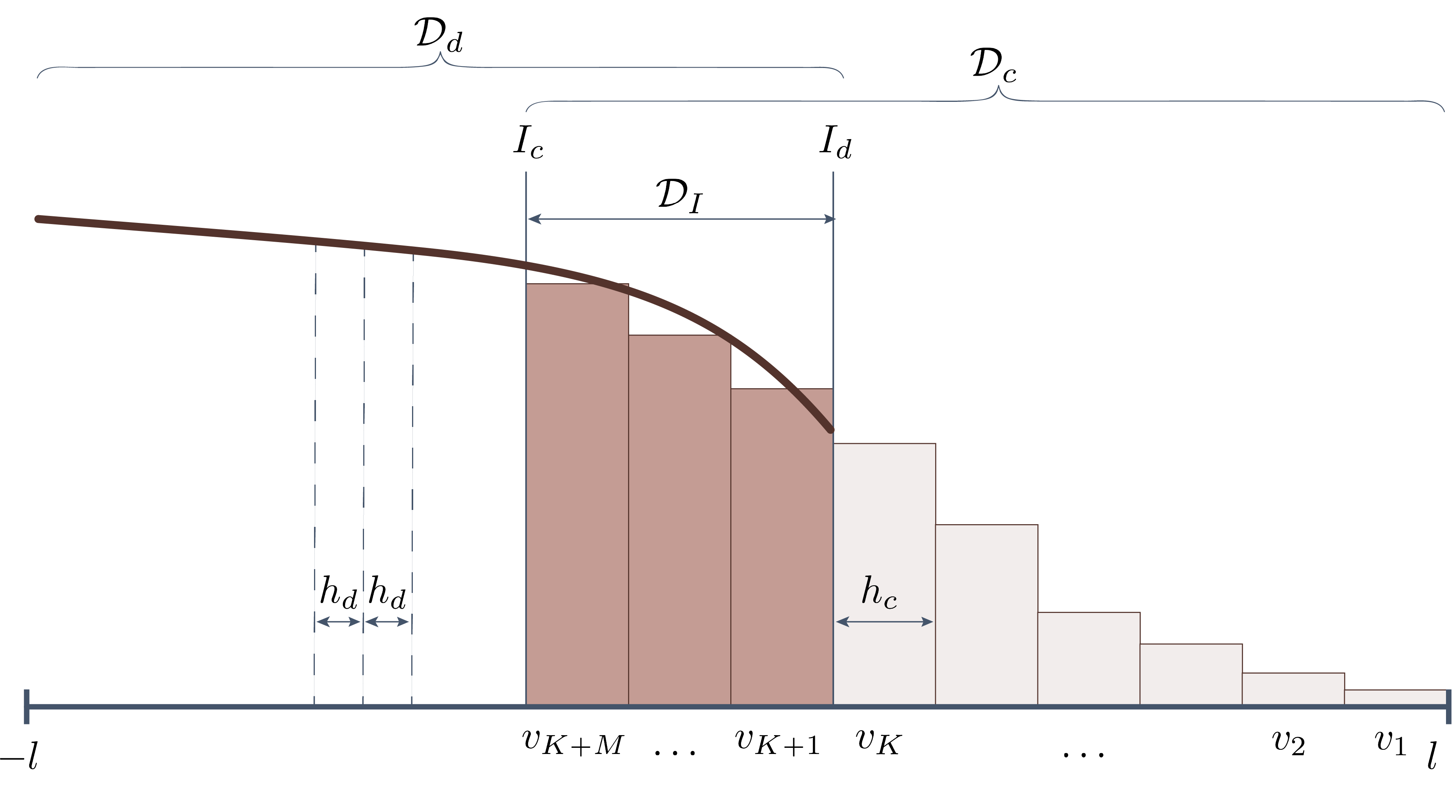}
} \hfill
\subfloat[]
{
\label{fig:HybridDomain_B}
\includegraphics[width=0.3842\linewidth]{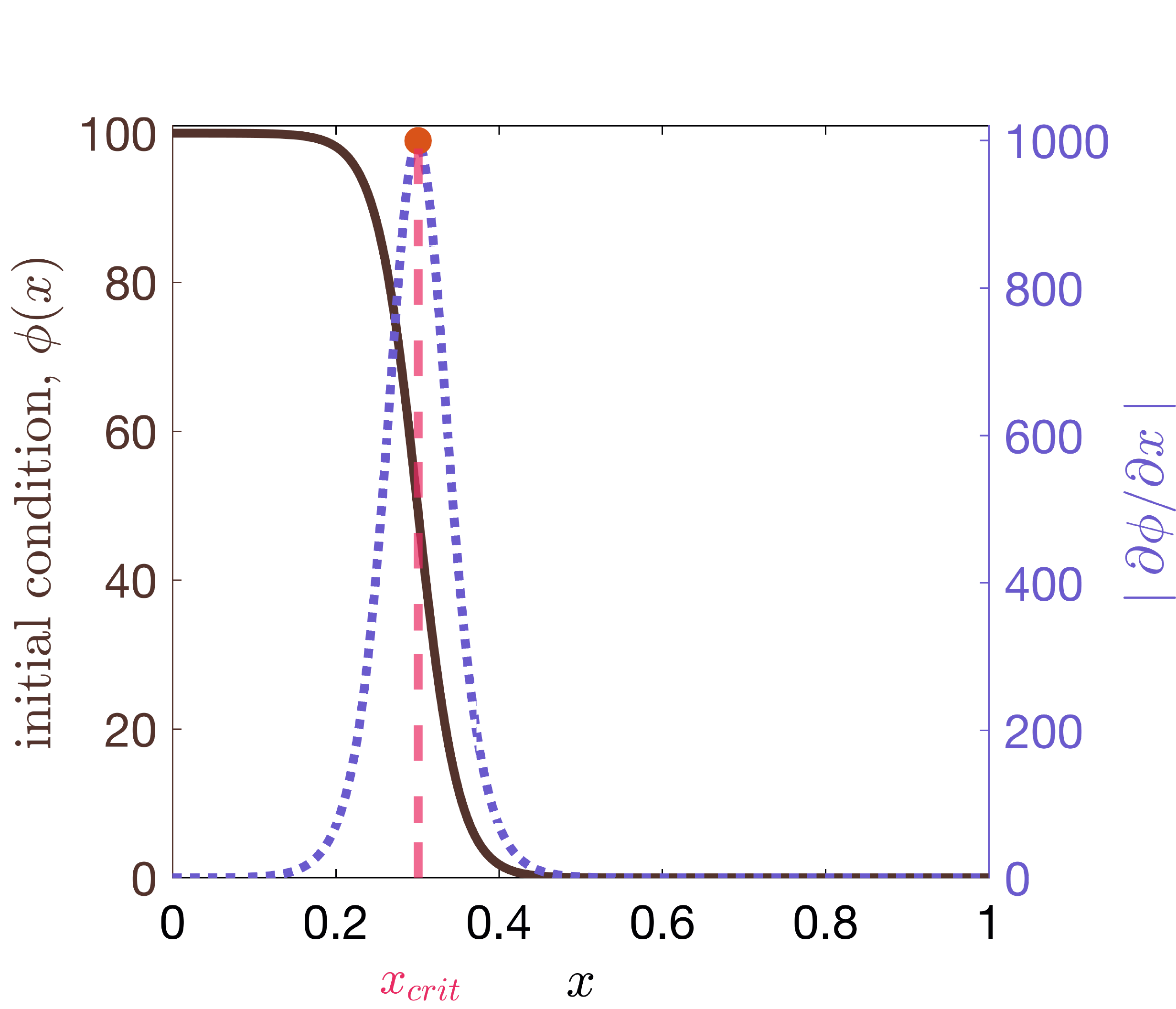}
}
\caption{\protect\subref{fig:HybridDomain_A} An illustration demonstrating the partitioning of the domain $\mathcal{D}=[-l,l]$ into the compartment-based subdomain, $\mathcal{D}_c$, simulated by the cRDME, and the deterministic subdomain, $\mathcal{D}_d$, described by the mean-field IPDE. The overlap region, $\mathcal{D}_I = [I_c,I_d] = \mathcal{D}_c \cap \mathcal{D}_d$, where both descriptions are valid, is determined by the positions of two interfaces, $I_c$ and $I_d$. Here, $h_c$ denotes the uniform voxel size, while $h_d$ is the mesh size for the finite difference discretisation in the deterministic subdomain. \protect\subref{fig:HybridDomain_B} An illustration of the critical point position, $x_{crit}$, determining the centre of the interface region at the initial time point. Here, $x_{crit}$ corresponds to the point at which the maximum of the absolute value of the gradient of the initial molecular distribution, $\phi(x)$, occurs.} \label{fig:HybridDomain}
\end{figure} 

The approach used here is based on the previous work, which introduced a hybrid method for systems exhibiting travelling wave behaviour \cite{harrison2016hybrid}. In particular, we assume that the simulation domain, $\mathcal{D}$, is divided into two overlapping subdomains (see Figure~\ref{fig:HybridDomain}~\subref{fig:HybridDomain_A}): one modelled using a stochastic compartment-based approach and the other using a macroscopic mean-field description. In what follows, we assume that the compartment-based description is valid in the right part of $\mathcal{D}$ (see Figure~\ref{fig:HybridDomain}~\subref{fig:HybridDomain_A}). The method can be similarly formulated if the positions of the subdomains are reversed.

The stochastic subdomain, $\mathcal{D}_c=[I_c,l]$, is partitioned into $K+M$ voxels of uniform size $h_c=(l-I_c)/(K+M)$. Out of these $K+M$ compartments, $M$ voxels lie within the interface region, $\mathcal{D}_I = [I_c, I_d]$. We denote by $N(k)$ the number of molecules in voxel $k$ for $k\in[1,K+M]$, while the corresponding concentration is given by $n_k=N(k)/(\Omega h_c)$. In the limit for the lattice size $h_c$ approaching zero, a continuous description, $n_c(x,t)$, valid within the compartment-based subdomain, can be obtained.

The macroscopic subdomain, $\mathcal{D}_d = [-l, I_d]$, is covered by a uniform mesh of size, $h_d$, for finite difference discretisation of the mean-field description, $n_d(x,t)$. In general, it is not necessary for the mesh points of $\mathcal{D}_d$ to coincide with the boundaries of the voxels (within the overlap region). The mesh size, $h_d$, must be chosen to ensure that the numerical schemes used to solve the mean-field limit IPDE provide accurate approximations.

\subsubsection{General Overview}

We now provide a general overview of the algorithm for our hybrid method. More detailed descriptions of the initial placement of interfaces, their dynamic update and coupling between the stochastic and mean-field regimes are discussed below.

\begin{enumerate}
\item Set parameter values. Set the initial time, $t=0$, and the time counter for updating the mean-field solution, $t_d=0$. 
\item Set the initial condition, $n(x,t=0) = \phi(x)$. We choose an initial condition such that the system evolves as a travelling wave.
\item Calculate the initial position of the interfaces (section~\ref{sec:init_interfaces}) and assign the corresponding subdomains for the compartment-based and mean-field regimes accordingly.
\item Set time step, $\tau$, as the waiting time to the next stochastic event using one of the algorithms for simulating stochastic systems (e.g. SSA or NSV method).
\item Update the state of the system within the compartment-based domain, $\mathcal{D}_c$, according to the chosen stochastic algorithm.
\item Update the time variable, $t = t + \tau$.
\item If $(t - t_d< \tau_d)$, where $ \tau_d$ is the time step for the mean-field discretisation, update the deterministic subdomain:
\begin{enumerate}
\item solve the mean-field IPDE in the time interval, $[t_d,t)$, within the deterministic subdomain, $\mathcal{D}_d$;
\item update the time counter, $t_d = t $;
\item if necessary, update the position of the overlap region (section~\ref{sec:update_interfaces}).
\end{enumerate}
\item Renormalise the number of molecules within the overlap region so that it is an integer. The renormalisation method used is explained in detail in \cite{spill2015hybrid}.
\item Iterate steps 4-9 until some stopping condition is satisfied. Here, we used the stopping criterion of reaching the final simulation time, i.e. $t\geq t_{final}$.
\end{enumerate}

\subsubsection{Initial Placement of the Interfaces} \label{sec:init_interfaces}

Here, we assume that a single pair of interfaces suffices to capture the spatial inhomogeneities of the molecule distribution. However, the method can be easily extended to multiple overlap regions. For each interface region, we need to specify (i) its centre and (ii) its width.

To determine the initial position of the centre of the overlap region, we identify a critical point, $x_{crit} \in \mathcal{D}$, corresponding to the maximum absolute value of the gradient of $\phi(x)$, where $\phi(x)$ represents the initial distribution of molecules. Figure~\ref{fig:HybridDomain}~\subref{fig:HybridDomain_B} illustrates the position of $x_{crit}$ for a sigmoidal initial condition. The algorithm can be extended to identify multiple critical points, including local extrema or all points where the absolute value of the gradient of $\phi(x)$ exceeds a certain threshold. In such cases, an overlap region must be defined for each critical point.

The interface region must be non-empty to ensure an accurate approximation of the system's dynamics \cite{harrison2016hybrid}. Additionally, the non-local nature of the bimolecular reactions in the convergent sFK system results in molecule fluxes between subdomains within the reaction radius, $R_a$. Therefore, the size of the overlap region must be at least $R_a$. In this study, we generally set the width of the interface region to $d = 2 R_a$ or $d = 4 R_a$.
 
Therefore, the positions of the two interfaces are calculated as follows: \\[-5pt]
$$I_c= \max\uf{-l, x_{crit}- 0.5 d} \text{ and } I_d=\min\uf{l, x_{crit} + 0.5 d}.$$

\subsubsection{Dynamic Adaptation of the Interface Region} \label{sec:update_interfaces}

To update the interface region, we formulate mutually exclusive criteria for extending one of the subdomains. We use the local concentration of molecules at the centre of the overlap region, $\mathcal{D}_I$, to determine whether the subdomains should be updated. As described above, the critical point defining the centre of the interface region is situated approximately at the centre of the wavefront. We introduce algorithm parameters, $0< r_1 < r_2 < 1$, to control the adaptation of the interfaces. Although both descriptions are valid in this region, the deterministic concentration of molecules is not subject to fluctuations. Thus, if the mean-field concentration of molecules $n_d(t,x_{crit})$ falls below $r_1 \mathcal{K}$ (indicating the wavefront has moved towards $I_c$), we extend $\mathcal{D}_c$ by adding a new voxel, with the number of molecules approximated by the deterministic concentration at the new voxel's centre. At the same time, the mean-field domain is reduced by the voxel width at the interface, $I_d$. Conversely, if the concentration $n_d(t,x_{crit})$ exceeds $r_2 \mathcal{K}$, we extend the domain $\mathcal{D}_d$ by replacing the voxel next to the interface $I_d$ with its mean-field description. The stochastic subdomain is then reduced by one compartment at the interface, $I_c$. The two interfaces and the centre point between them must then be updated accordingly. If changes in the size of the compartment-based domain affect the stochastic algorithm (e.g. if the NSV method is used), it must also be updated to ensure its correct performance. Typically, in our simulations, we use $r_1 \in [0.4, 0.6]$ and $r_2 \in [0.8, 1)$. 

Since we use the mean-field molecule density as the criterion for updating the interface region, domain adaptation is performed at most once every deterministic time step, $\tau_d$, which is typically much larger than the time step for more frequent stochastic reactions, $\tau$. This ensures that the interface update algorithm does not significantly increase the computational complexity of the method.

\subsubsection{Coupling the Mean-Field and the Stochastic Descriptions at the Overlap Region}

The coupling algorithm of our hybrid method must ensure consistent molecule concentration across the interfaces. The non-local nature of the convergent formulation of the sFK system (see section~\ref{subsec:Fisher_cRDME}) implies that stochastic and deterministic densities influence each other through molecular fluxes from reactions occurring within the reaction radius, $R_a$, beyond the subdomain borders. This direct coupling between the two descriptions must be accounted for in our hybrid method.

Following the approach proposed in \cite{harrison2016hybrid}, we balance molecular fluxes at the interfaces, $I_c$ and $I_d$. This is achieved by calculating the additional flux of molecules entering the stochastic subdomain through $I_c$ for voxels within the reaction radius, $R_a$, from this interface. For the mean-field density, we prescribe an additional annihilation flux coming from the stochastic subdomain and set an appropriate Dirichlet boundary condition at the interface, $I_d$. As before, we assume that the compartment-based domain is on the right.

\textbf{Coupling for the Mean-Field Description ~~~~} Within the overlap region, the evolution of the deterministic mean-field molecular concentration, $n_d(x,t)$, corresponding to the convergent sFK system, is governed by the following IPDE:
\begin{align}
&\frac{\partial n_d}{\partial t} (t,x) = D \frac{\partial^2 n_d}{\partial x^2}(t,x) + \lambda n_d(x,t) \nonumber \\
& ~~~~~~~~~~~~~~~~ - \underbrace{\mu n_d(t,x) \af{ \displaystyle \int_{\mathcal{D}_d} b(x-y) n_d(t,y) dy + \mathcal{A}(t,x) }}_{\mathcal{A}_d(t,x)},
\label{eq:IPDEHybrid} \\
&n_d(t,I_d)=\bar n, \qquad \qquad \frac{\partial n_d}{\partial x}(t,-l)=0. &&
\label{eq:BCHybrid}
\end{align}
\noindent Here, we note the Dirichlet boundary condition, $\bar{n}$, at the interface, $I_d$, and the additional annihilation flux $\mathcal{A}(t,x)$ coming from the compartment-based subdomain (compare with Eq.~\eqref{eq:NLFK}). The expression for the former was derived in \cite{harrison2016hybrid} and reads:
\begin{equation}
\bar n := \frac{N(K)+N(K+1)}{2 \Omega h_c} = \frac{n_K+n_{K+1}}{2}.
\label{eq:BoundaryCondition}
\end{equation}
\noindent We recall that voxel $K$ (respectively $K+1$) lies to the right (respectively left) of the interface, $I_d$ (see Figure~\ref{fig:HybridDomain}~\subref{fig:HybridDomain_A}). It can be shown \cite{harrison2016hybrid} that the boundary condition given by Eq.~\eqref{eq:BoundaryCondition} ensures the consistency of the molecular density across $I_d$, i.e. $n_d(t,I_d)=n_c(t,I_d) + \mathcal{O}(h_c^2) \approx n_c(t,I_d)$.

\begin{figure}[t!]
\subfloat[]
{
\label{fig:IndexSets_A}
\includegraphics[width=0.4522\linewidth]{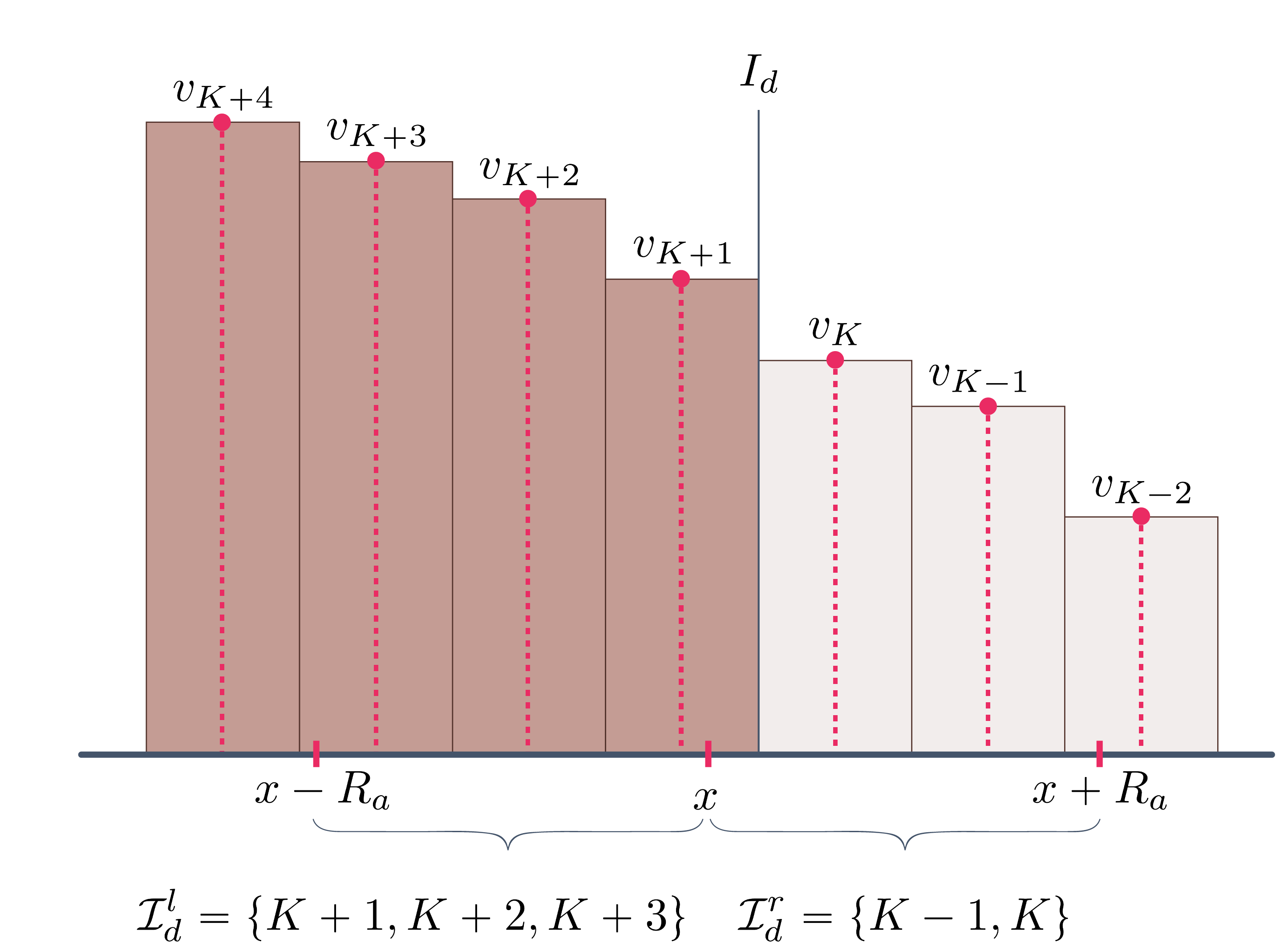}
} \hfill
\subfloat[]
{
\label{fig:IndexSets_B}
\includegraphics[width=0.4916\linewidth]{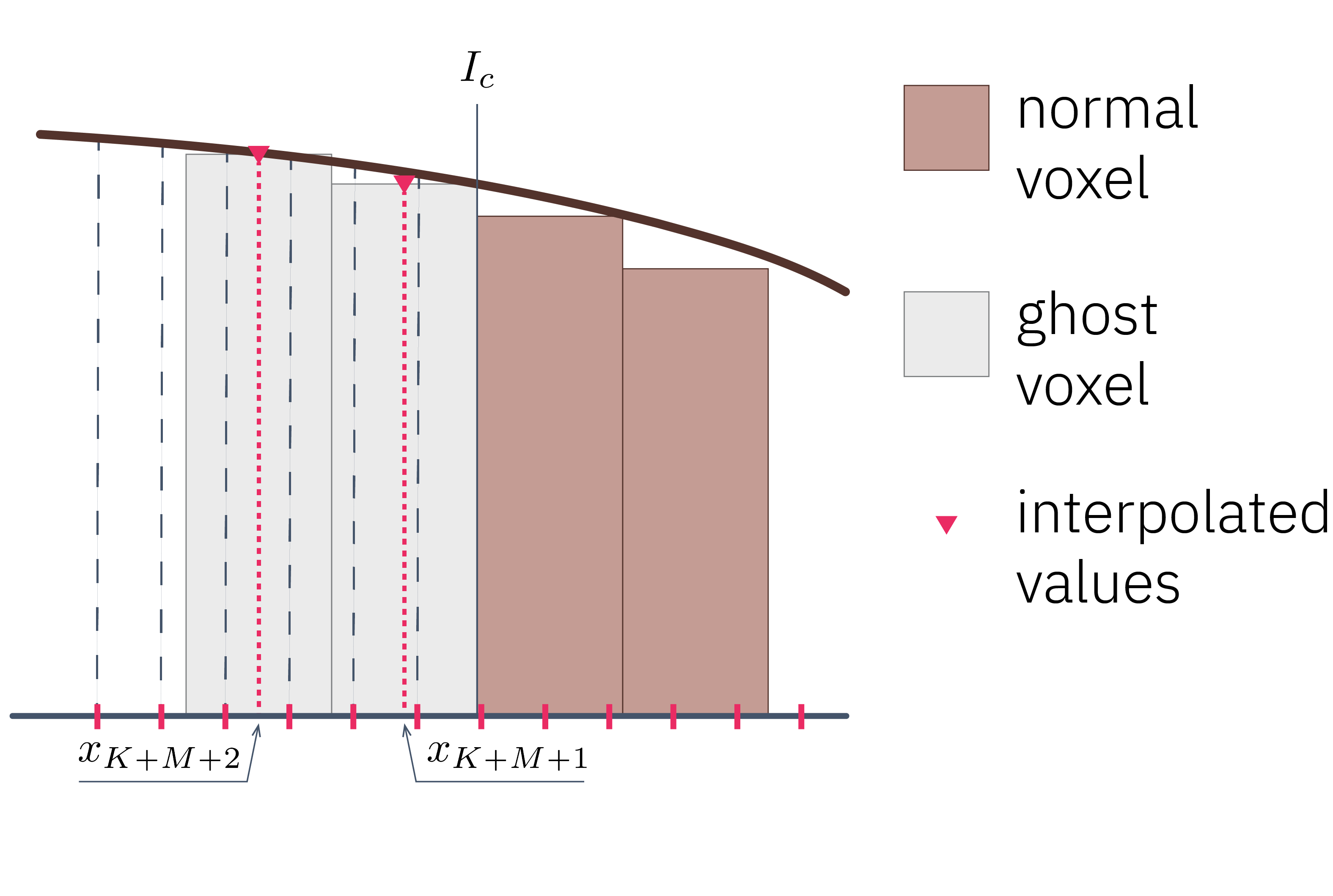}
}
\caption{\protect\subref{fig:IndexSets_A} We introduce index sets, $\mathcal{I}_i^{l}(x)$ and $\mathcal{I}_i^{r}(x)$, of voxels which lie within the reaction radius, $R_a$, from point $x$ to the left and right of the interface $I_i$, respectively. Here, $i \in \uf{c,d}$ is the interface label. Specifically, left-lying sets of voxels are defined as $\mathcal{I}_i^{l}(x) \coloneqq \uf{k ~|~ (l-(k-0.5)h_c) \in [x-R_a, I_i) }$, while the indices of the voxels to the right of the interface $I_i$ are given by $\mathcal{I}_i^{r}(x) \coloneqq \uf{k ~| ~(l-(k-0.5)h_c) \in  [I_i, x+R_a]) }$. We note that these index sets can include ghost voxels (see \protect\subref{fig:IndexSets_B}). The illustration shows these index sets for the interface, $I_d$. \protect\subref{fig:IndexSets_B} An illustration of linear interpolation for ghost voxels $(K+M+1)$ and $(K+M+2)$.} \label{fig:IndexSets}
\end{figure}

We further introduce the annihilation flux, $\mathcal{A}(t,x)$ (see Eq.~\eqref{eq:IPDEHybrid}), which arises due to the non-local reaction kinetics to ensure convergence of the RDME. It is calculated as follows
\begin{equation}
\mathcal{A}(t,x) =
\begin{cases}
\frac{1}{2 R_a} \displaystyle \sum_{k\in \mathcal{I}_d^{r}(x) \cap \mathcal{I}_d^{r}(I_d)} p(x,k) \frac{N(t,k)}{ \Omega}, \quad & \text{if } I_d -R_a < x \leq I_d, \\
0, & \text{otherwise,}
\end{cases}
\label{eq:AnnihilationFlux}
\end{equation}
\noindent where $p(x,k) = |v_{k} \cap B_{R_a}(x)|/|v_{k}|$ is the continuous version of $p(k,m)$ from Eq.~\eqref{eq:pkm} and the index set, $I_d^r(\cdot)$, is as described in Figure~\ref{fig:IndexSets}~\subref{fig:IndexSets_A}.

It is straightforward to check that in the limit $h_c\rightarrow 0$, the annihilation fluxes through the interface $I_d$, $\mathcal{A}_d(t,x)$ and $\mathcal{A}_c(t,x)$, for stochastic and deterministic descriptions, respectively, are equal. Here, $\mathcal{A}_d(t,x)$ is outlined in Eq.~\eqref{eq:IPDEHybrid} with $\mathcal{A}(t,x)$ given by Eq.~\eqref{eq:AnnihilationFlux}, whereas the stochastic flux due to annihilation events is given by the following expression:
\begin{equation}\label{eq:StochFluxI0}
\mathcal{A}_c(t,x_k) = \frac{\mu}{2 R_a} \frac{1}{\Omega h_c } N(t,k) \displaystyle \sum_{m \in \mathcal{I}_d^{ l}(x_k) \cup \mathcal{I}_d^{r}(x_k)} p(k,m) \frac{N(t,m)}{\Omega}.
\end{equation}
\noindent Here, the index set, $\mathcal{I}_d^{r}(x)$, is defined in Figure~\ref{fig:IndexSets}~\subref{fig:IndexSets_A}.

\textbf{Coupling for the Compartment-Based Regime ~~~~} At the interface, $I_c$, compartments lying to the right within the reaction radius $R_a$ are influenced by the molecular flux from the deterministic subdomain. Specifically, for all voxels $v_k$ in $\mathcal{D}_c$ with $k \in \mathcal{I}_c^{r}(I_c)$ (see Figure~\ref{fig:IndexSets}~\subref{fig:IndexSets_A}), we must account for the additional flux, $\psi^a_k(t, \Delta t)$, due to annihilation reactions between molecules from different subdomains occurring at time $t$ during the time step $\Delta t$. To achieve this, we introduce ghost voxels to the left of the interface $I_c$ (assuming the stochastic subdomain is on the right). The number of ghost voxels corresponds to the number of indices in the set $\mathcal{I}_c^{r}(I_c)$. The number of molecules in the ghost voxels is determined by interpolating the molecular concentration at the voxel centre using the two nearest points from the $n_d(t,x)$ discretisation. Figure~\ref{fig:IndexSets}~\subref{fig:IndexSets_B} illustrates this process. Consequently, the additional annihilation flux into voxel $v_k$ for $k \in \mathcal{I}_c^{r}(I_c)$ is given by the following expression:
\begin{equation}
\psi_{k}^a (t, \Delta t)= \Delta t \skpp \mu \frac{h_c}{2 R_a} n_c(t,x_k) \displaystyle \sum_{ m \in \mathcal{I}_c^{l}(x_k) \cap \mathcal{I}_c^{l}(I_c)} p(k,m) \skp \widetilde{n}_{m}.
\label{eq:psi_a}
\end{equation}
\noindent Here, $\widetilde{n}_{m}$ denotes the concentration of molecules in ghost voxel $m$, and $x_k$ is the center of the $k$-th voxel. The summation in Eq.~\eqref{eq:psi_a} encompasses all voxels within the reaction radius from $x_k$ that are not part of the stochastic subdomain (i.e., ghost voxels, $\widetilde{n}_{m}$). This equation, therefore, accounts solely for the molecular flux originating from the deterministic subdomain. The stochastic dynamics within compartments $v_k \subset \mathcal{D}_c$, including annihilation events, are defined as described previously in section~\ref{subsec:Fisher_cRDME}.

In the limit as the lattice size, $h_c$, approaches zero, it can be shown that the annihilation fluxes, $\mathcal{A}_c(t,x_k)$ and $\mathcal{A}_d(t,x_k)$, for $x_k \in [I_c,I_c + R_a]$ are identical:
\begin{align*}
& \mu \frac{h_c}{2 R_a} n_c \of{t,x_k} \displaystyle \sum_{m \skp \in \skp \mathcal{I}_c^{l}(x_k) \skp \cup \skp  \mathcal{I}_{c}^{r}(x_k) } p (k,m) ~ n_c \of{t,x_m} \\
& \stacksymbol{=}{h_c \rightarrow 0}{} \mu \frac{1}{2 R_a} n_d \of{t,x_k} \displaystyle \int_{x_k-R_a}^{x_k+R_a} n_d(t,y) dy.
\end{align*}
\noindent To verify this, we note that within the overlap region, $n_d(t,x)$ serves as a good approximation of $n_c(t,x)$. Therefore, for any $x \in [I_c,I_c + R_a]$, $n_d(t,x) \approx n_c(t,x) $. Additionally, to the left of the interface, $I_c$, we used the deterministic molecule concentration for the linear interpolation of $n_c(t,x)$ in ghost voxels. Thus, these fluxes are identical, and the presence of the interface, $I_c$, does not affect the molecular concentration.
  
For voxel $(K+M)$, adjacent to the interface $I_c$ (see Figure~\ref{fig:HybridDomain}~\subref{fig:HybridDomain_A}), incoming molecules from the deterministic domain due to diffusion must be considered in addition to the annihilation flux. The diffusion flux through the interface, $I_c$, was calculated in previous work \cite{harrison2016hybrid} and is given as follows:
\begin{equation}
\psi_{K+M}^d (t, \Delta t) = \Delta t \frac{D}{h_c^2}\left(\widetilde{n}_{K+M+1} - n_{K+M}\right).
\label{eq:psiD}
\end{equation}

Summarising, for voxel $(K+M)$ at the interface $I_c$, we must correct its molecular concentration by incorporating molecular fluxes from both annihilation events and diffusion (Eqs.~\eqref{eq:psi_a}-\eqref{eq:psiD}). For all other voxels $k \neq K+M$ influenced by the deterministic subdomain, we adjust their molecular concentrations solely based on the flux from annihilation using Eq.~\eqref{eq:psi_a}. This ensures the consistency of molecular concentration across the $I_c$ interface.

\subsection{Simulation Results} \label{subsec:Results_cRDME}

\begin{figure}[h]
\centering
\includegraphics[width=\linewidth]{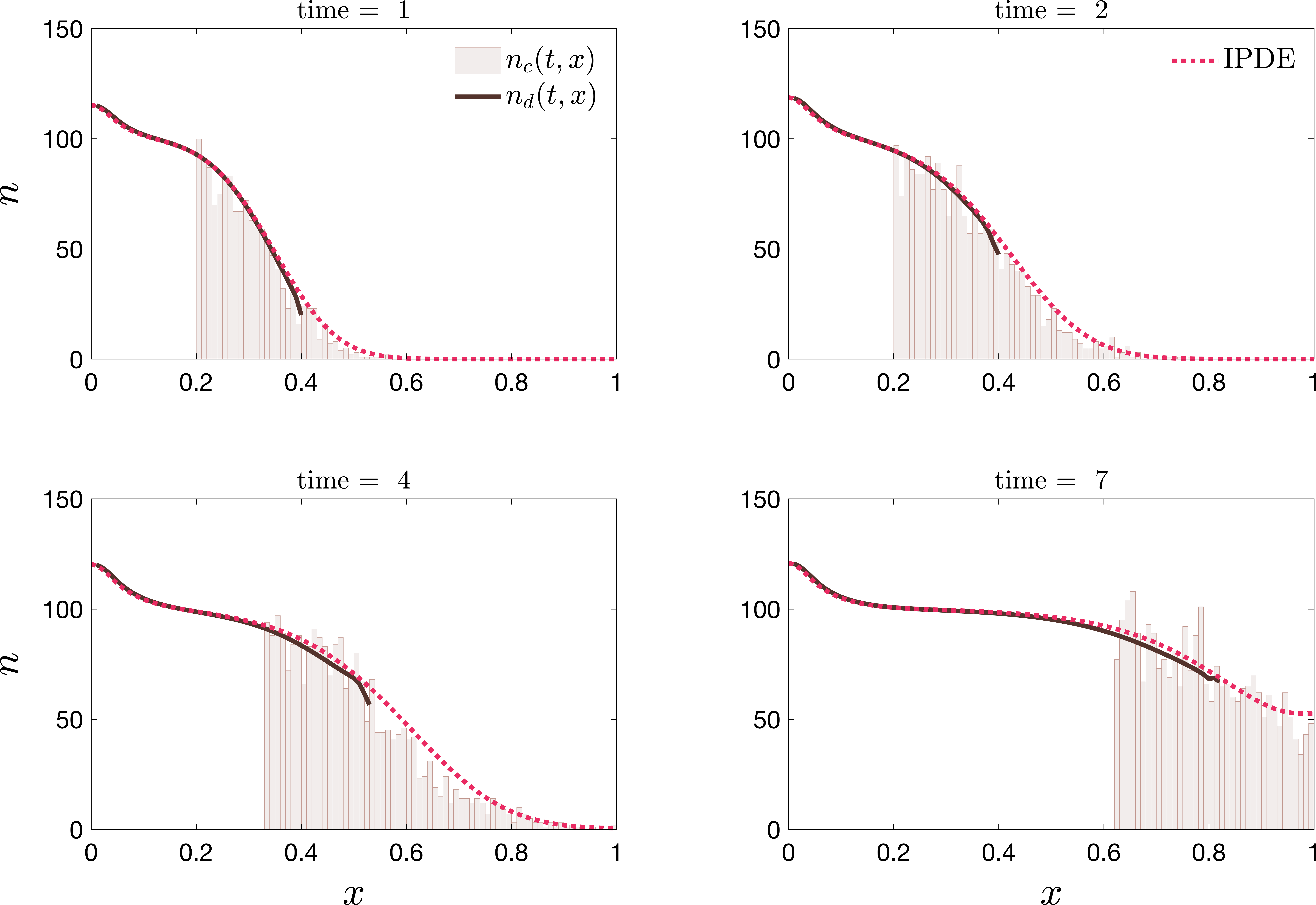} 
\caption{Four snapshots of a numerical simulation of our hybrid method for the non-local Fisher-Kolmogorov system. Dashed lines indicate the solution of the deterministic IPDE given by Eq.~\eqref{eq:NLFK} at the same time moment. The width of the interface region was set to $d = 4 R_a$ and the system size to $\Omega = 100$. The remaining parameters are listed in Table~\ref{tab:params2}. The initial molecular concentration used for this simulation is given by a smooth sigmoidal function, $\phi(x) = 0.5 \skpp \mathcal{K} (1 - \tanh(20(x-0.3)))$.}
\label{fig:HybridStepMoments}
\end{figure}

We illustrate our hybrid method for the cRDME with characteristic simulation results for the sFK model. The parameter values used in these simulations are listed in Table~\ref{tab:params2}. Figure~\ref{fig:HybridStepMoments} displays four snapshots of the wavefront propagation over time. We note that, in these simulations, the noise parameter in the stochastic regime was set to a high value to demonstrate the robustness of our hybrid method (corresponding system size parameter, $\Omega = 100$). The small hump forming at the domain border is due to the no-flux boundary condition and the non-locality of annihilation reactions.

\begin{table}[b!]
\centering
\begin{tabular}{| l | l | p{2.8cm} || l | p{2.8cm} | p{2.8cm} |}
\hline
Param.& Value & Description & Param.& Value & Description \\
\hline
$h_c=h_d$ & $0.01$ & Lattice spacing. & $D$ & $4e-3$ (Figure~\ref{fig:HybridStepMoments}); ~~~~~ $4e-5$ (Figure~\ref{fig:SpeedComparison})& Diffusion coefficient. \\
$\lambda$ & $1.0$& Birth rate. & $\mathcal{K}$ & $100$ & Carrying capacity. \\
$R_a$ & $0.05$ & Interaction radius for bimolecular reactions. & $t_{final}$ & $7$ (Figure~\ref{fig:HybridStepMoments}); \phantom{word ~~~~~~} $30$ (Figure~\ref{fig:SpeedComparison}) & Simulation time.\\
$r_1$ &$0.5$ & Interface adaptation parameter. & $r_2$ &$0.8$ & Interface adaptation \phantom{w} parameter.\\ \hline
\end{tabular}
\caption{Parameters used in numerical simulations of the convergent Fisher-Kolmogorov system.}
\label{tab:params2}
\end{table}

\subsubsection{Comparison}

Although Figure~\ref{fig:HybridStepMoments} showcases a single simulation, a slight delay in wave propagation speeds between the hybrid algorithm and the purely deterministic IPDE can already be observed. In the final snapshot at $t=7$, the IPDE front is notably advancing faster. This discrepancy arises because the wavefront advancement in our hybrid method is governed by the stochastic regime, which typically exhibits slower wave propagation speeds that depend on the noise level and approach the deterministic mean-field estimate as the noise tends to zero (equivalently, system size in stochastic simulations tends to infinity). The mean-field estimation for the front advancement in the non-local Fisher-Kolmogorov system (Eq.~\eqref{eq:NLFK}) is not affected by the integral term and is given by $c_{d} \approx 2~\sqrt[]{D \lambda}$ for an infinite domain \cite{gourley2000travelling}. In order to evaluate the performance of our hybrid method, in this section, we compare wave propagation speeds in our hybrid method, purely stochastic simulations using the NSV method, and the mean-field limit estimate, $c_{d}$.

In order to obtain estimates of the wave propagation speed for stochastic and hybrid simulations, we employ the method introduced in \cite{robinson2014adaptive}. This method relies on the linear temporal growth of the (averaged) total molecular mass in a system with travelling wave behaviour, where the slope of its growth corresponds to the wavefront speed. We denote by $\mathcal{M}_s(t)$ and $\mathcal{M}_h(t)$ the total masses of the system at the time point $t$ in purely stochastic and hybrid simulations, respectively. Then, the wave speeds can be approximated by the following expressions:
\begin{equation}
c_{s} \approx \frac{\langle \mathcal{M}_s (t_2) \rangle - \langle \mathcal{M}_s (t_1) \rangle}{t_2 - t_1}, \qquad \qquad  c_{h} \approx \frac{\langle \mathcal{M}_h (t_2) \rangle - \langle \mathcal{M}_h (t_1) \rangle}{t_2 - t_1},
\label{eq:SpeedFormula}
\end{equation}
\noindent where $\langle \cdot \rangle$ denotes the mean taken over multiple stochastic realisations, and the total molecular masses are given by: 
\begin{equation}
\mathcal{M}_s(t) = \frac{1}{\Omega} \displaystyle \sum_{k=-K+1}^{K} N(k), \qquad \mathcal{M}_h(t) = \frac{1}{\Omega} \displaystyle \sum_{k=1}^{K+M} N(k) + \displaystyle \int_{\mathcal{D}_d \setminus \mathcal{D}_I} n_d(t,x) dx.
\end{equation}

\begin{figure}[t!]
\centering
\includegraphics[width=\linewidth]{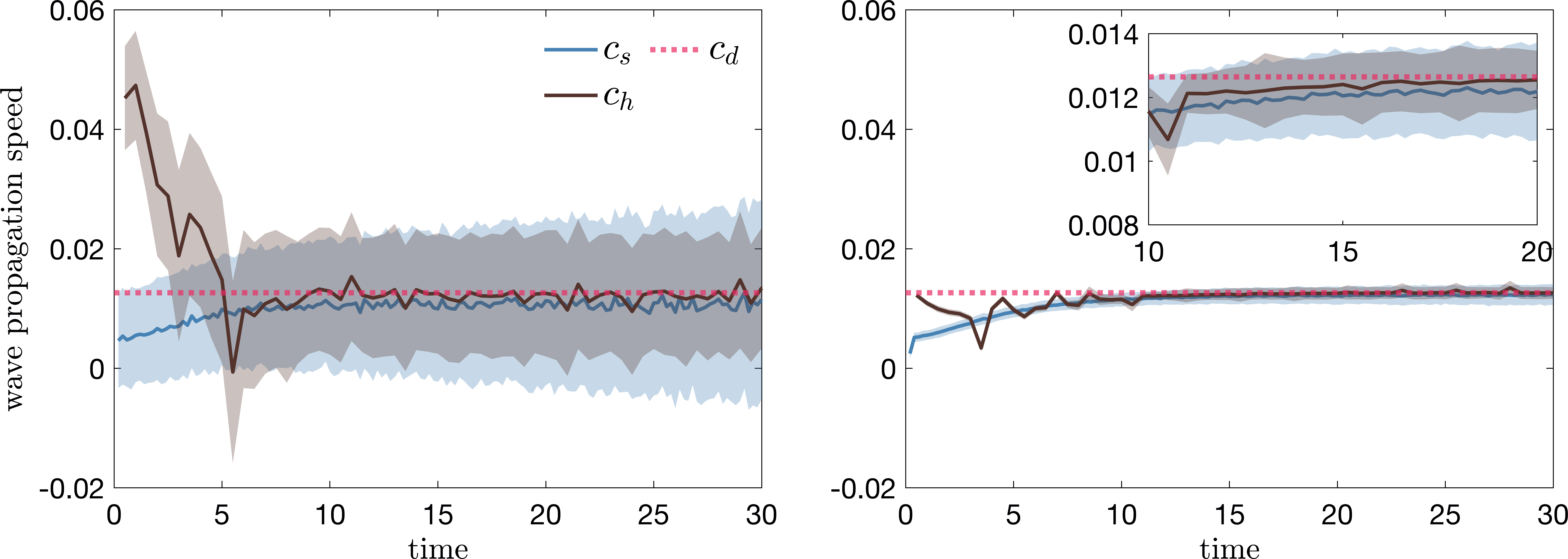} 
\caption{Wave propagation speeds for the convergent Fisher-Kolmogorov system estimated from purely stochastic ($c_s$, in blue), hybrid ($c_h$, in dark brown) simulations, with system sizes, $\Omega = 100$ (left panel) and $\Omega = 1000$ (right panel), and the theoretical mean-field estimate, $c_d$. Solid lines indicate the mean values averaged over 500 realisations, while shaded regions show their standard deviation. Here, the width of the interface region was set to $d = 2 R_a$ and the initial condition, $\phi(x)$, is as in Figure~\ref{fig:HybridStepMoments}. The remaining parameters are listed in Table~\ref{tab:params2}.}
\label{fig:SpeedComparison}
\end{figure}

Figure~\ref{fig:SpeedComparison} presents the results of the wavefront propagation estimation for two system sizes: $\Omega = 100$ (left panel) and $\Omega = 1000$ (right panel). After an initial transient phase ($t \lessapprox 10$), the system stabilises into a travelling wave propagating at constant average speeds, $c_s$ and $c_h$, for the stochastic and hybrid simulations, respectively. As the noise level decreases (right panel), the estimates converge more closely to the deterministic limiting speed, $c_d$. The average wave propagation speeds are $c_s \approx 0.010676$ and $c_h \approx 0.012132$ for $\Omega = 100$, and $c_s \approx 0.012131$ and $c_h \approx 0.012596$ for $\Omega = 1000$. The macroscopic limit estimate for the parameters used in these simulations is $c_d = 0.012649$. As expected, the purely stochastic simulations show the slowest wave propagation speed, while the hybrid method's estimate lies between the purely stochastic and mean-field theoretical estimates.

We also recorded the computational times for performing these simulations using stochastic and hybrid approaches. For an individual realisation, the mean time to run a stochastic simulation with the NSV implementation was $20.9306 \pm 0.8807$ seconds. This time was reduced by approximately 48-fold for hybrid simulations, which required around $0.43309 \pm 0.0333$ seconds per realisation. These estimates were obtained for a system size of $\Omega = 100$. All simulations were performed on a standard personal PC. While simulation times for the relatively simple sFK model are manageable even for the purely stochastic approach, the reduction in simulation time provided by the hybrid method can offer significant advantages for more complex models.

In summary, hybrid methods for convergent RDME require the overlap region width to be on the order of the interaction radius for multi-molecular reactions. While this adds to the computational complexity compared to the traditional RDME approach, hybrid methods offer significant reductions in simulation time compared to purely stochastic methods. This introductory section demonstrates the formulation of such a hybrid method for the convergent Fisher-Kolmogorov system. However, more complex chemical reaction networks can also be implemented within this framework.

\section{Conclusion \& Discussion}

%####################
This chapter reviewed hybrid models of travelling waves in reaction-diffusion systems, emphasizing their relevance to biological processes such as animal and cell movement. We discussed the broader context of deterministic pulled fronts and highlighted the differences between various wave types. Future work will focus on extending hybrid methods to bistable waves and other complex systems, enhancing our ability to model biological invasions accurately.
%##############
 
Hybrid methodologies have been extensively shown to be powerful tools for effective simulation of stochastic reaction-diffusion systems. In this chapter, we have reviewed the different methodologies and formulations proposed in the literature (see section~\ref{sec:Intro}), as well as our recent work devoted to addressing several limitations in the previous implementations. In particular, in section~\ref{sec:age}, we have reviewed our recent work on extending the hybrid methodology to multiscale models of tumour growth. Stochastic multiscale models account for fluctuations in both the number of cells (the usual demographic noise present in reaction-diffusion systems) and the intracellular cell dynamics. The situation analysed in section~\ref{sec:age} illustrates the importance of properly considering the latter. 

In section~\ref{sec:Multiple_species}, we have summarised our extension of the hybrid method to account for multi-species reaction-diffusion systems (e.g. Lotka-Volterra systems). Since these systems are known to produce heterogeneous spatial patterns, one must tackle the emergence of several interfaces \cite{spill2015hybrid}. Finally, in section~\ref{sec:hybrid_cRDME}, we proposed a hybrid method that allows us to simulate the convergent version of the reaction-diffusion Master Equation. This method involves developing an approach to address non-local reaction mechanisms. 

Beyond the extensions discussed in this chapter, there is still plenty of room for further development and applications of hybrid methods. A recent example is the paper by \cite{liang2023dipsvirus}, where they studied the stochastic effects of the interaction between viruses and particles lacking replication genes that can interfere with wild-type viruses and serve as potential therapeutics in a two-dimensional model of infection. Extension of the current algorithms to higher spatial dimensions is an essential step to extend the current remit of hybrid methods regarding their applications. 

Fronts are observed in systems beyond the ones we have described in detail in this review, and it would be a subject of interest to extend hybrid methodologies to such systems, for example in the contexts of waves and fronts in bistable and excitable systems.

Another avenue of potential extensions of the current hybrid schemes is growing and compliant domains \cite{smith2021growth}. The mechanical microenvironment has revealed itself to be a key factor in many biological and biomedical problems from developmental biology to tumour growth, and therefore should be accounted for in order to further extend the range of application of hybrid methods.

\section*{Acknowledgements}

D.S. and T.A. would like to thank the Isaac Newton Institute for Mathematical Sciences, Cambridge, for support and hospitality during the programme \emph{Mathematics of movement: an interdisciplinary approach to mutual challenges in animal ecology and cell biology}, supported by EPSRC grant EP/R014604/1. This work has been funded by the Spanish Research Agency (AEI), through the Severo Ochoa and Maria de Maeztu Program for Centers and Units of Excellence in R\&D (CEX2020-001084-M). D.S. and T.A.  thank CERCA Program/Generalitat de Catalunya for institutional support. D.S. and T.A. have been funded by grant PID2021-127896OB-I00 funded by MCIN/AEI/10.13039/501100011033 ‘ERDF A way of making Europe’. J.C. has been partially supported by Grant C-EXP-265-UGR23 funded by Consejería de Universidad, Investigación e Innovación \& ERDF/EU Andalusia Program, by Grant PID2022-137228OB-I00 funded by the Spanish Ministerio de Ciencia, Innovación y Universidades, MICIU/AEI/10.13039/501100011033 \& “ERDF/EU A way of making Europe”, and by  Grant QUAL21-011 (Modeling Nature) Consejería de Universidad, Investigacion e Innovacion of the Junta de Andalucía. N.B. and P.G have been supported by Grant PID2022-141802NB-I00 (BASIC) the research network RED2022-134573-T funded by MCIN/AEI/ 10.13039/501100011033 and, by ‘ERDF A way of making Europe’.

%\input{references}
%\bibliographystyle{}
%\bibliography{references.bib}
%\printbibliography

% \input{author/references_review}
\end{document}